\documentclass[12pt,preprint]{aastex}
\usepackage{amssymb, amsmath}
\usepackage{color}
\usepackage{morefloats}
\usepackage{hyperref}
\usepackage{natbib}

\shortauthors{Neugent et al.}

\begin{document}

\title{The Luminosity Function of Red Supergiants in M31}

\author{Kathryn F.\ Neugent\altaffilmark{1,2}, Philip Massey\altaffilmark{2,3}, Cyril Georgy\altaffilmark{4}, Maria R.\ Drout\altaffilmark{5,6}, Michael Mommert\altaffilmark{2}, Emily M.\ Levesque\altaffilmark{1}, Georges Meynet\altaffilmark{4}, and Sylvia Ekstr\"{o}m\altaffilmark{4}}

\altaffiltext{1}{Department of Astronomy, University of Washington, Seattle, WA, 98195}
\altaffiltext{2}{Lowell Observatory, 1400 W Mars Hill Road, Flagstaff, AZ 86001; kneugent@lowell.edu}
\altaffiltext{3}{Department of Astronomy and Planetary Science, Northern Arizona University, Flagstaff, AZ, 86011-6010}
\altaffiltext{4}{Department of Astronomy, University of Geneva, CH-1290 Versoix, Switzerland}
\altaffiltext{5}{Department of Astronomy and Astrophysics, University of Toronto, 50 St. George Street, Toronto, Ontario, M5S 3H4 Canada}
\altaffiltext{6}{The Observatories of the Carnegie Institution for Science, 813 Santa Barbara St., Pasadena, CA 91101}

\begin{abstract}
The mass-loss rates of red supergiant stars (RSGs) are poorly constrained by direct measurements, and yet the subsequent evolution of these stars depends critically on how much mass is lost during the RSG phase. In 2012 the Geneva evolutionary group updated their mass-loss prescription for RSGs with the result that a 20$M_\odot$ star now loses 10$\times$ more mass during the RSG phase than in the older models. Thus, higher mass RSGs evolve back through a second yellow supergiant phase rather than exploding as Type II-P supernovae, in accord with recent observations (the so-called ``RSG Problem"). Still, even much larger mass-loss rates during the RSG phase cannot be ruled out by direct measurements of their current dust-production rates, as  such mass-loss is episodic. Here we test the models by deriving a luminosity function for RSGs in the nearby spiral galaxy M31 which is sensitive to the {\it total} mass loss during the RSG phase. We carefully separate RSGs from asymptotic giant branch stars in the color-magnitude diagram following the recent method exploited by Yang and collaborators in their Small Magellanic Cloud studies. Comparing our resulting luminosity function to that predicted by the evolutionary models shows that the new prescription for RSG mass-loss does an excellent job of matching the observations, and we can readily rule out significantly larger values.
\end{abstract}
\keywords{galaxies: individual (M31) --- stars: massive --- stars: mass-loss --- supergiants}

\section{Introduction}
Red supergiants (RSGs) begin their lives on the main sequence as 8-30$M_\odot$ O and B-type stars before burning through their core hydrogen and evolving across the HR-Diagram (HRD). They briefly (for $\sim$10,000-100,000 years) pass through the yellow supergiant (YSG) phase as they cool down to temperatures below 4100K and expand to radii of  100-1500$\times$ that of the sun. During this time they fuse helium in their core and sustain luminosities that are $\sim$10,000-100,000 times greater than that of our Sun. Currently the majority of such stars are thought to end their lives by dying as Type II-P supernovae (see, e.g., \citealt{Meynet2015}).

The evolution and final fate of RSGs is heavily influenced by their mass-loss rates which are quite large (as high as $10^{-4}M_\odot$ yr$^{-1}$ compared to the Sun's mass loss rate of $10^{-14}M_\odot$ yr$^{-1}$) and episodic. \cite{Ekstrom2012} significantly revised the mass-loss prescription for RSGs for the Geneva evolutionary models, based upon the newer measurements of  \cite{Sylvester1998} and \cite{vanLoon1999}, as discussed in \cite{Crowther2001}. The differences with the older \cite{deJager1988} relation are small, except for the highest luminosity RSGs. They found that for RSGs above some certain mass limit of 15$M_\odot$, the outer layers of the star would exceed the Eddington luminosity, and in this situation increased the mass-loss rate by an additional factor of 3 for the duration of the event. The net result of these changes is that the time-averaged mass-loss rate of a 20$M_\odot$ RSG at solar metallicity was a factor of 10 greater than that obtained by the previously adopted \cite{deJager1988} relationship traditionally used in the older evolutionary models (see, e.g., \citealt{Schaller1992}).    The implications of this change were many and varied, and include the result that the mass limit for stars to becoming Wolf-Rayets (WRs) is decreased and more stars go through a second YSG phase after becoming RSGs rather than exploding directly as supernova.  As suggested by \cite{Ekstrom2012}, this later difference could explain the so-called ``RSG problem," the lack of observed high-luminosity RSG progenitors to Type II-P SNe \citep{Smartt2009}. The older Geneva evolutionary models had done a poor job in predicting the relative number of YSGs as a function of luminosities \citep{Drout2009,Neugent2010}, but after this revision to the RSG mass-loss rates (as well as many other improvements), the newer models reproduced the observations very well \citep{Drout2012,Neugent2012}.  However, until now there has been no other observational tests of the revised mass-loss rate prescription during the RSG phase. 

This revision in the RSG mass-loss rate was profound for the higher mass RSGs, but the direct measurements of the mass-loss rates do not preclude even larger changes \citep{Meynet2015}, and there is no direct constraint based upon stellar atmosphere modeling. Because these stars have such low surface gravities ($\log g \sim 0$) and escape velocities ($< 100$ km s$^{-1}$), Stan Owocki (private communication, 2006) has argued that it's a little like walking across the room with a glass of water filled to the brim: even a small jiggle can lead to dramatic increases in the mass loss for a RSG.  Mass-loss rates have only been directly measured for around 50 stars and the results are highly inconsistent \citep{Mauron2011, vanLoon2005, Beasor2016} (see \citealt{Meynet2015}, Figure 1 for a comparison to the Geneva model mass-loss rate prescriptions). In addition, the mass-loss rates themselves are uncertain: to measure the mass-loss rate of individual RSGs, all we can do is measure the dust production rate and then infer a total mass-loss rate by multiplying by an uncertain large number (100-500$\times$) depending on what we assume for the gas-to-dust ratio.  And even this idea of a ``standard" mass-loss rate is questionable. For one, what we measure today might not relate to the time-averaged mass-loss rate. For instance, \citet{Smith2001} demonstrated that VY CMa, an extremely luminous RSG, underwent a period of enhanced mass loss 1000 years ago \citep{Decin2006}. It is unknown how often these outbursts occur and what fraction of the measured mass-loss rates might be caused by an outburst. \citet{Meynet2015} and \citet{Georgy2015} have argued that the actual time-averaged mass-loss rates could be a factor of 10 or even 25 times higher than what is currently adopted.

\begin{figure}
\epsscale{0.75}
\plotone{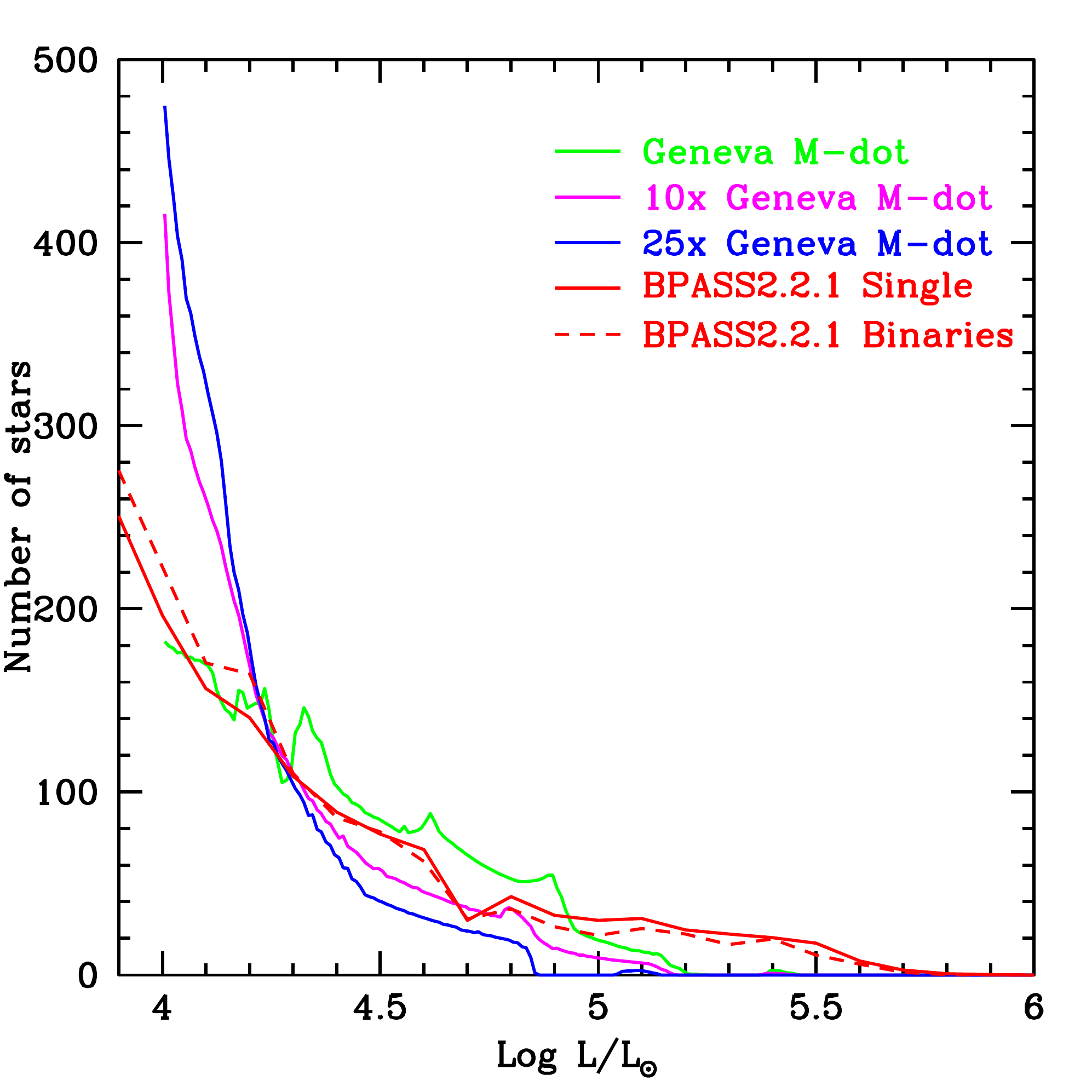}
\caption{\label{fig:mdot} The dependence of the RSG luminosity function on the assumed mass-loss rate. The theoretical predictions have all been scaled to a sample of 1000 RSGs with $\log L/L_\odot \ge 4.0$ and analyzed with 0.1~dex bins, similar to the sample we analyze here.  The green line comes from the Geneva mass-loss prescriptions described in \cite{Ekstrom2012}, which are in accord with more modern estimates of the observed rates as described in the text. The magenta and blue lines show the luminosity functions for enhancing this rate by a factor of 10 and 25, respectively.  All of these Geneva predictions were computed using {\sc syclist} \citep{2014A&A...566A..21G}. We also include the luminosity functions (red)  from the BPASS 2.2.1 models \citep{Stanway2018} both for single stars (solid) and with binaries (dashed); we are grateful to J. J. Eldridge for providing these predictions.   Although the effect of binaries is to reduce the number of RSGs by a factor of $\sim$2 for a given star-formation rate, it has almost no effect on the {\it shape} of the luminosity function.  Note in particular the good agreement between the Geneva and BPASS 2.2.1 predictions.}
\end{figure}

Here we follow up on the suggestion of \citet{Meynet2015} and \citet{Georgy2015} of using the RSG luminosity function as an observational test of the time-averaged RSG mass-loss rates.    As RSGs evolve and go through core helium fusion, their effective temperatures barely change but their luminosity increases as the helium core grows. Increased mass-loss rates (whether as single or binary stars) shorten the RSG lifetimes of these stars by removing their outer layers since as more mass is removed, the envelope to core ratio will change. Once the core represents a certain fraction of the star, the star will evolve back blue-ward. So, the more mass lost, the faster the blue-ward evolution will occur. The more luminous RSGs will lose more mass than the less luminous RSGs because mass-loss rates are luminosity-dependent. Thus, as mass loss rates increase, the ratio between the number of higher luminosity RSGs to the number of lower luminosity RSGs will decrease. This is shown in Figure~\ref{fig:mdot}. As the luminosity increases, the relative number of RSGs decrease in all the models, due to the decreasing number of higher mass progenitors (all assume a Salpeter IMF). For a given number of total stars, the Geneva high mass-loss models ($10\times$ and $20\times$ models) predict far more low luminosity RSGs, and fewer high luminosity RSGs \citep{Geneva, Ekstrom2012}. The BPASS 2.2.1 binary and single-star models \citep{BPASS22} predict a very similar luminosity function as the standard Geneva model, except at the higher luminosity end, where they predict significantly more stars\footnote{We are indebted to J.J. Eldridge for kindly computing the expected RSG luminosity distributions for us from her latest BPASS models.}. The BPASS 2.2.1 models adopt the \cite{deJager1988} prescription for mass-loss during the RSG phase, although of course the binary versions also include mass-loss driven by the companions \citep{BPASS1,BPASS2,Stanway2018}.

\section{The Need for New Near-Infrared Data}
\label{Sec-NeedNIR}
Obtaining a useful luminosity function of RSGs has three requirements. First we must compute luminosities at the 0.05~dex level for a sample down to $\log L/L_\odot\sim4.0$ in order to create a well-sampled histogram. Fainter than this runs the risk of contamination of faint red foreground stars as shown in Sec.~\ref{Sec-AGBs}. Requiring luminosities at a 0.05~dex precision level ruled out using RSGs in the Milky Way due to uncertain distances in the pre-{\it Gaia} era. Additionally even in the post-{\it Gaia} Data Release 2 (DR2) era, it is messy and uncertain to identify a volume-limited sample in the presence of large and variable reddenings. This luminosity requirement drove us to the near-infrared (NIR) rather than the optical, as detailed below. Secondly, as our goal is not only to obtain an observationally robust luminosity function, but to compare to current model predictions for mass loss rates, we currently require stars that have metallicities of solar and above ($1-1.8 Z_\odot$), ruling out using RSGs in the LMC or Small Magellanic Cloud (SMC) as the lower metallicity single-star Geneva Evolutionary grids do not produce high luminosity RSGs. Thirdly, we need a large enough sample (several hundred stars) so the luminosity function is not dominated by stochastic, small-number statistics at the bright end. These three considerations lead us to M31's RSG population, as the galaxy is nearby with a well-determined distance (760~kpc, \citealt{vandenbergh2000} and references therein), low reddening ($A_V$$\sim$1.0~mag for the RSGs, \citealt{MasseySilva,Massey2016}), and a suitable metallicity that is about 1.5$\times$ solar \citep{Sanders}.

The NIR is preferable to the optical for three main reasons. First, we are less sensitive to the reddening than in the optical; the extinction at {\it K$_s$} is only 12\% of what it is at $V$ ($A_K = 0.12 A_V$; see, e.g., \citealt{Schlegel1998}), meaning that uncertainties in the extinction have a much smaller effect on the derived luminosity\footnote{For convenience, we have included the various relations we've adopted or derived throughout this paper in Table~\ref{tab:FunFacts}.}.  Similarly, $J-K$ colors are less affected by reddening than are optical colors; while $E(B-V)=A_V/3.1$, $E(J-K)=A_V/5.8$ \citep{Schlegel1998}.  This is particularly an issue for RSGs, as they suffer from circumstellar extinction due to their dust production. The typical OB star in M31 has $A_V$ of 0.5~mag \citep{LGGSII}, while the RSGs in our sample typically have an average $A_V$ of 1.0~mag \citep{MasseySilva}; we see a similar effect when we compare Galactic RSGs to OB stars in the same clusters; see \citet{Smoke}.  (We will delve into the reddening issue more exactly in Sec.~\ref{Sec-transforms}.)

Second, the bolometric corrections (BCs) at {\it K} are much less sensitive to the determination of effective temperatures than at $V$. The effective temperatures of RSGs are $\sim$3500-4100~K \citep{Levesque2005}; as a result, their flux $F_\lambda$ peaks in the far red (7000-8000\AA). Over this temperature range BC$_V$ varies from $-2.2$ to $-0.8$ mag (i.e., by 1.4~mags), while BC$_K$ varies from +2.9 to +2.5 (i.e., by 0.4~mag); see \citet{Levesque2005}.   At the same time, $(J-K)_0$ is quite sensitive to the effective temperature, changing from 1.20 (3500~K) to 0.86 (4100~K), a difference $\Delta (J-K)_0 = 0.34$, according to the MARCS models used by \citet{MasseySilva}.  For comparison, $(V-R)_0$ changes by almost the same amount, from 1.09 to 0.76 ($\Delta (V-R)_0 = 0.33$), although it is more affected by reddening. Thus for the same photometric precision, the uncertainty in the BC will be 3.5$\times$ smaller using NIR colors than in the optical, even in the absence of reddening, due to the smaller change in the BC over a similar range in color and temperature.

The third factor driving us to the NIR is that RSGs are photometrically more variable at $V$ than at {\it K}. \citet{Levesque2006} found that the typical $V$-band variability during a year was 0.9~mag, while {\it K}-band variability amongst RSGs is $\sim$0.15~mag \citep{Josselin,MasseySilva}. This irregular variability is due to several causes, including variable dust production (producing increased circumstellar extinction), and the existence of large convective zones on the surface of these stars that have high contrast between the hotter and cooler spots in {\it V} than at {\it K}.

\begin{figure}
\epsscale{1.0}
\plotone{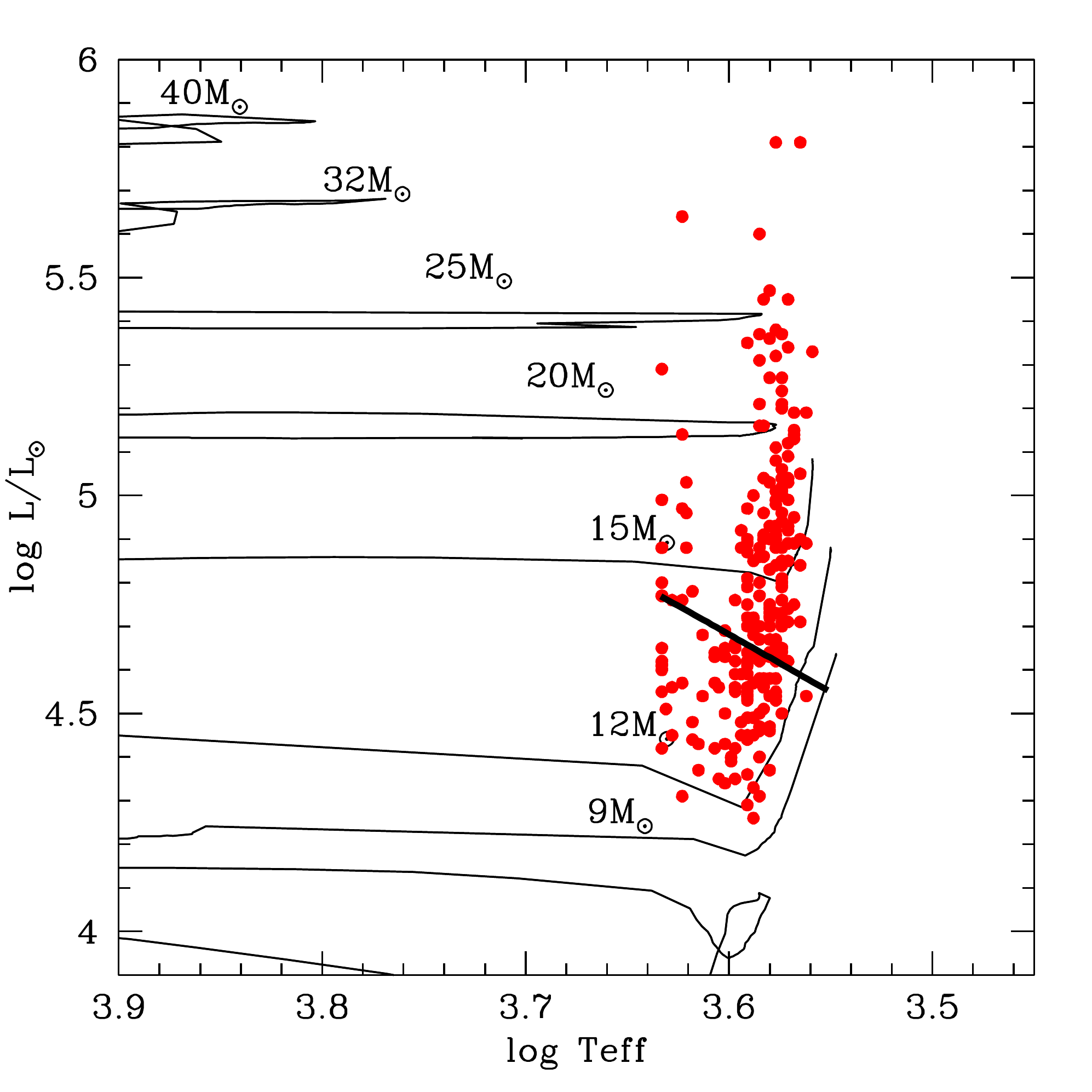}
\caption{\label{fig:Evans}H-R Diagram of red supergiants in M31 based on 2MASS data. The evolutionary tracks of \citet{Ekstrom2012} for $z=0.014$ are shown along with the (initial) masses. The black line near $\log L/L_\odot \sim 4.8$ denotes the completeness limit of the 2MASS K-band photometry.  Adapted from \cite{Massey2016}.}
\end{figure}

\citet{Massey2016} use the Two Micron All Sky Survey (2MASS) photometry to construct the H-R diagram shown in Figure~\ref{fig:Evans}. As above, accurate {\it J-K} colors and {\it K-}band magnitudes are needed to determine good bolometric luminosities. The 2MASS photometry is complete (at 10$\sigma$) only to $K=15.0$ for the 6x catalog \citep{2Mass6x}, which included coverage of M31. We indicate this completeness limit by the black band. This completeness is purely photometric, and ignores issues of crowding, which causes increasing losses at fainter magnitudes in crowded fields such as found in M31; this is especially true given the 2\arcsec\ resolution of 2MASS. Thus we are currently limited to $\log L/L_\odot$$\sim$4.8 by the lack of good NIR photometry, while from Figure~\ref{fig:Evans} we see we really would like to go down to $\log L/L_\odot$ of $\sim$4.0 to include $\sim 9 M_\odot$, say\footnote{The astute reader will also notice some mis-match between the tracks and the locations of the points in Figure~\ref{fig:Evans}; in our analysis here we will use an improved relation between temperature and color, and also treat the reddening in a better manner than did \cite{Massey2016}.} . This corresponds to $K$$\sim$16.2-16.7 for a cool (3500~K) and warm (4100~K) RSG, respectively. 

Given the reasons detailed above, we set out to obtain new near-IR photometry of RSGs within M31. 

\section{Observations and Reductions}
\label{Sec-obs}
All the {\it J} and {\it K} photometry comes from new images taken with the Wide Field Camera (WFCAM) on the 3.8-meter United Kingdom Infrared Telescope (UKIRT) located on Mauna Kea, Hawai'i. The data were obtained as part of the UKIRT Service Program (Proposal U/17B/UA03); UKIRT is owned by the University of Hawai'i (UH) and operated by the UH Institute for Astronomy; operations are enabled through the cooperation of the East Asian Observatory.  Observing time was granted through the Steward Observatory Telescope Allocation Committee.  

\begin{figure}
\epsscale{1.0}
\plotone{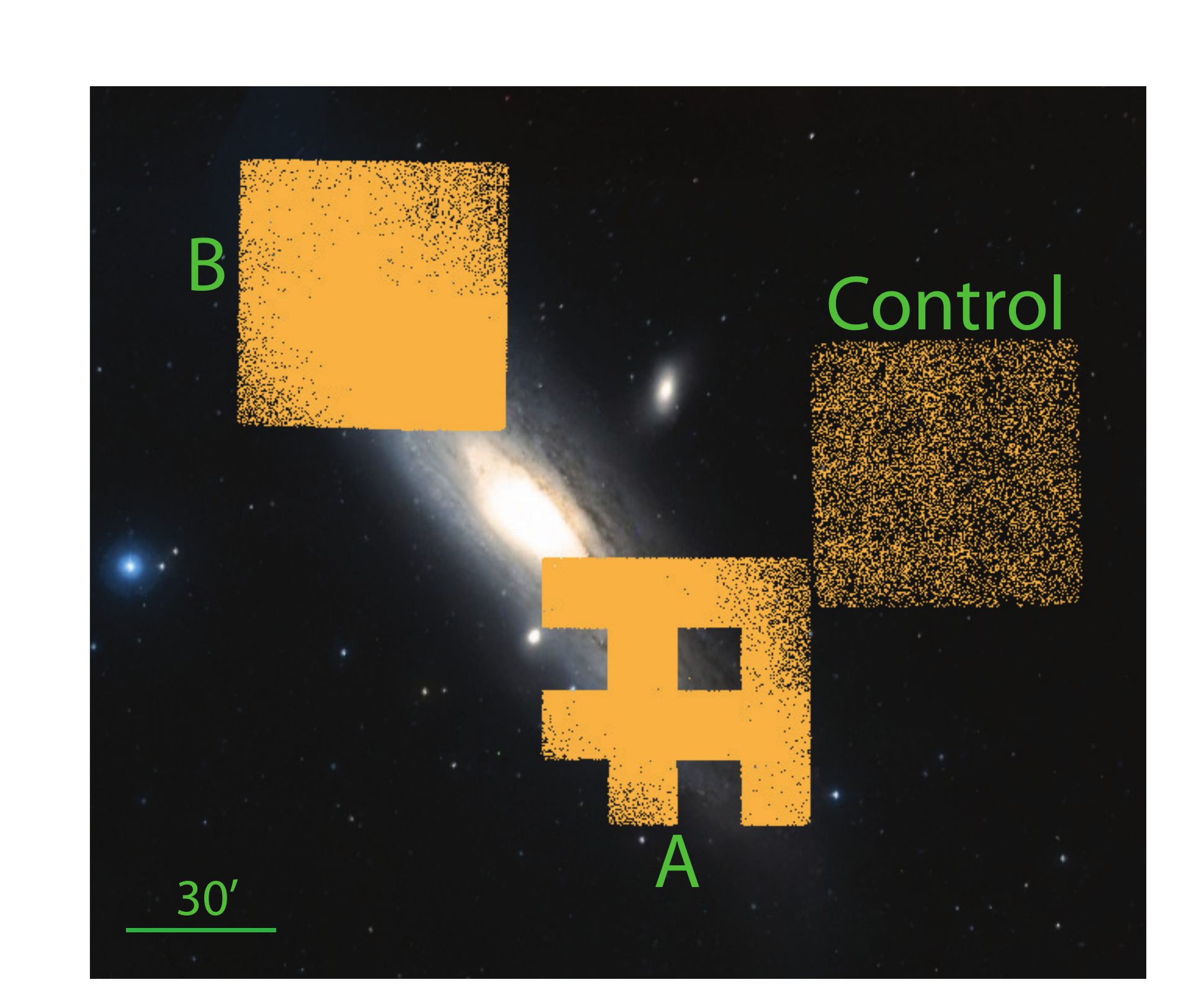}
\caption{\label{fig:where} Location of our fields in M31. The three UKIRT fields (A, B, and control) are shown (see Table~\ref{tab:where}) by indicating all of the stars with photometry as yellow points. The non-square shape of field A is due to the planned dithers not being executed properly.}
\end{figure}

We selected two M31 fields and one nearby ``control" field\footnote{As we discuss in Section~\ref{Sec-gaia}, our control field was unfortunately a tad {\it too} ``nearby."} as shown in Figure~\ref{fig:where} and Table~\ref{tab:where}. The scientific requirement was to achieve an uncertainty of 0.05~dex in $\log L$, or roughly 0.1~mag, at $\log L/L_\odot$$\sim$4.0. We aimed for a S/N of 30 at $K$=17.0 and $J$=18.0, about 0.5~mag deeper than needed to assure completeness. This gives us an error of $<$0.05~mag in $J-K$ for our faintest stars, leading to an uncertainty of $<$100~K in the effective temperature scale (similar to what we achieve by fitting models to our optical spectrophotometry; see \citealt{Levesque2005}), or an error of 0.08~mag in the BC at K. The reddening at $V$ varies by 0.5~mag at the faint end (\citealt{MasseySilva} and discussed in depth below in Sec.~\ref{Sec-transforms}); at {\it K} this introduces another 0.06~mag uncertainty.  Add these together with the uncertainty of 0.03~mag in the $K$-magnitude itself, and we achieve an uncertainty of 0.10~mag, or 0.04~dex in $\log L$. It is not clear what effect the $\sim$0.15~mag variation in {\it K} has. Perhaps it is compensated for changes in temperatures. But, even if we add this in, we achieve an error of 0.07~dex in $\log L$, not much worse than our 0.05~dex goal.

To make our observations more compatible with UKIRT's queue scheduling, we split each of the three fields into two ``visits," which were separated by days or weeks as shown in Table~\ref{tab:where}. WFCAM consists of four Rockwell Hawaii-II (HgCdTe 2048$\times$2048) arrays, each covering 13.65\arcmin\ on the sky, and separated by 12.83\arcmin\ gaps. The scale is 0\farcs4 per pixel. (For further details, see \citealt{WFCAM}.)  For each visit, we planned a sequence of exposures starting with a set of 10 individual 10-sec exposures in {\it J}. The individual exposures were co-added in order to build up the signal-to-noise. The process was then repeated in {\it K$_s$}, using 36 individual 10-sec exposures. The telescope was then offset by 795\arcsec\ (13.25\arcmin) and the process executed three more times in order to fill in the gaps between the chip. The entire sequence was then executed again during the second visit on a different night. Had everything gone according to plan, each field would have been observed twice in {\it J} and {\it K$_s$} covering a 51\arcmin\ $\times$ 51\arcmin\ area with some overlap between the dithers. However, there was no motion between dither position 1 and dither position 2 on the first visit to Field A, resulting in a 25\% loss of areal coverage for that field. The second visit was executed as planned.
 
The images were processed by the Cambridge Astronomical Survey Unit (CASU) and made available via the WFCAM Science Archive, part of the Wide Field Astronomy Unit hosted by the Institute for Astronomy, Royal Observatory, Edinburgh. An overview is given by \citet{WFCAMArchive}, with complete details given at the CASU web site\footnote{http://casu.ast.cam.ac.uk/surveys-projects/wfcam}. The data products include the calibrated chip-by-chip (``detector frame") stacked images and the corresponding chip-by-chip source catalogs with coordinates, instrumental magnitudes, aperture corrections, and photometric zero-points. The latter are based upon 2MASS stars observed throughout the night. Thanks to these expert pipeline products, our task was primarily book-keeping rather than calibration or photometry. Our process was to first transfer the FITS catalog tables into something more usable, and produce calibrated photometry for each source adopting the pipeline's ``aperture 3" values, which correspond to the ``core" flux, defined as a radius of 5 pixels (2.0\arcsec). According to the source catalogs, the typical (median) seeing was 4.0 pixels (1\farcs6), and ranged from 3.1 pixels (1\farcs2) to 6.5 pixels (2\farcs6). Correspondingly the typical (median) aperture correction was $-0.21$~mag, ranging from $-0.14$~mag to $-0.48$~mag. 

For each field we then had 64 catalogs: 2 visits $\times$ 2 filters $\times$ 4 chips $\times$ 4 dithers. Our next task was to match detections in the {\it J} and {\it K$_s$} exposures for each chip/visit/dither, keeping only stars that were detected in both filters. That reduced the burden to 32 catalogs containing $K_s$ and $J-K_s$ photometry for each chip/dither/visit. For each visit and dither we then combined the photometry of the four chips to produce 8 source catalogs of photometry. We then combined the data of the four dithers, averaging the photometry for stars in common in the regions of overlap. That resulted in two source catalogs, one for each visit.  Finally, we insisted that stars be detected in each of the two visits (using a 0\farcs5 matching criterion), producing a single catalog of objects for each field, with the photometry from the two visits averaged.  (Note that this eliminated one-quarter of the coverage we would otherwise have had for field A, as one of the dithers was missing.)  Thus, each star was observed at least twice in both {\it J} and {\it K$_s$}.  

Checking our combined photometry against the 2MASS values showed that there was invariably very little difference in the $K_s$-band values, but typically a +0.025~mag difference in the $J$-band photometry, in the sense that the 2MASS value minus UKIRT. We restricted the comparison to the standard 2MASS point source catalog (not the x6x version) and for stars with J and K$_s$ values between 12.0 and 14.0 and quality ratings of ``AAA."  Fainter than this, the scatter in the 2MASS data were too large; brighter than 12 we found that the UKIRT magnitudes showed larger and larger differences for increasingly brighter sources, characteristic of a non-linearity or saturation effect. We thus removed any UKIRT data for stars with $J\leq12$ or $K\leq12$ (after correcting for the median difference compared to 2MASS), and supplemented the catalog with 2MASS sources at the brighter end.  Care was taken for the 2MASS additions to Field A to make sure they fell in the same areas as the other stars were observed.

\begin{figure}
\epsscale{0.48}
\plotone{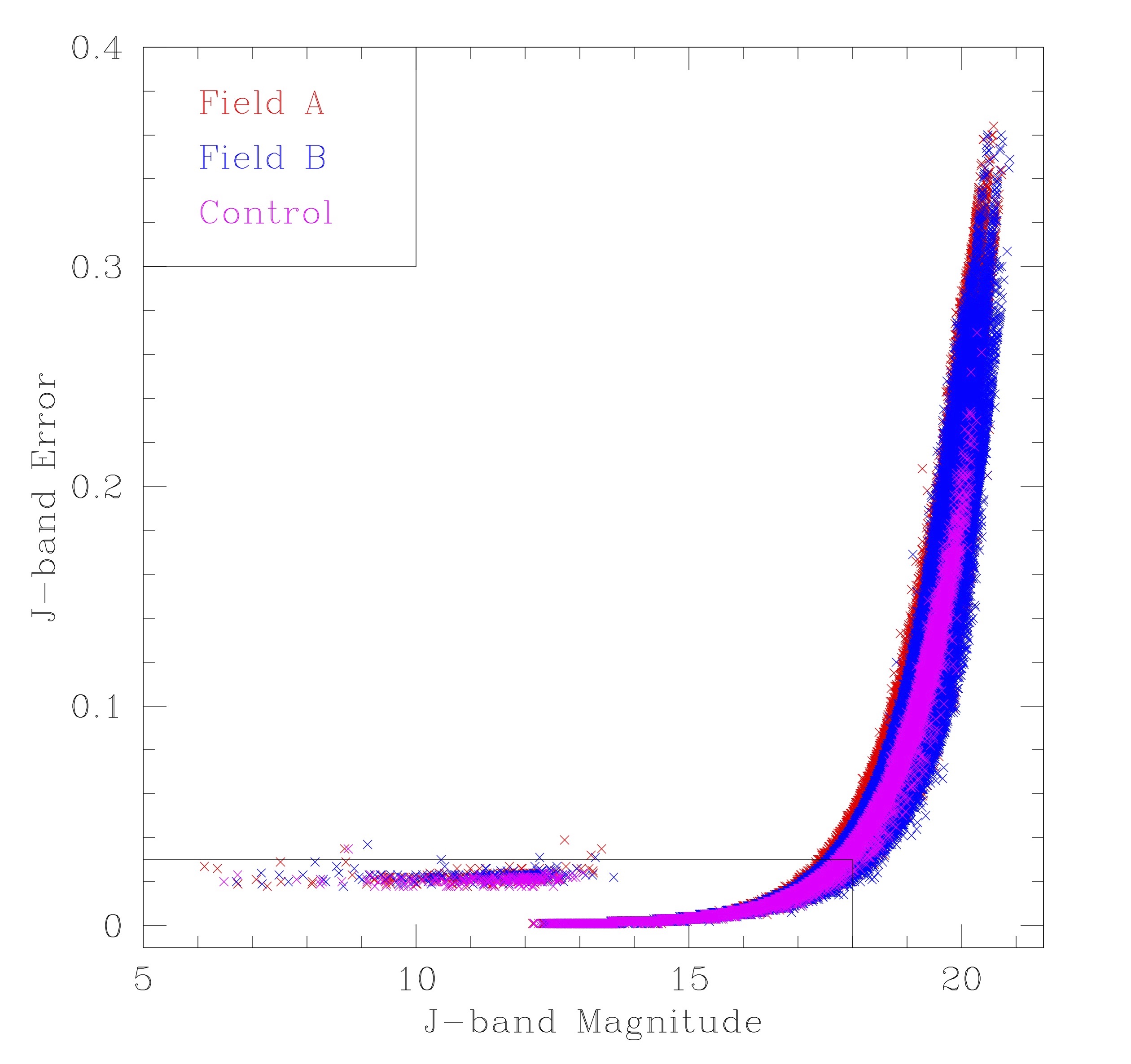}
\plotone{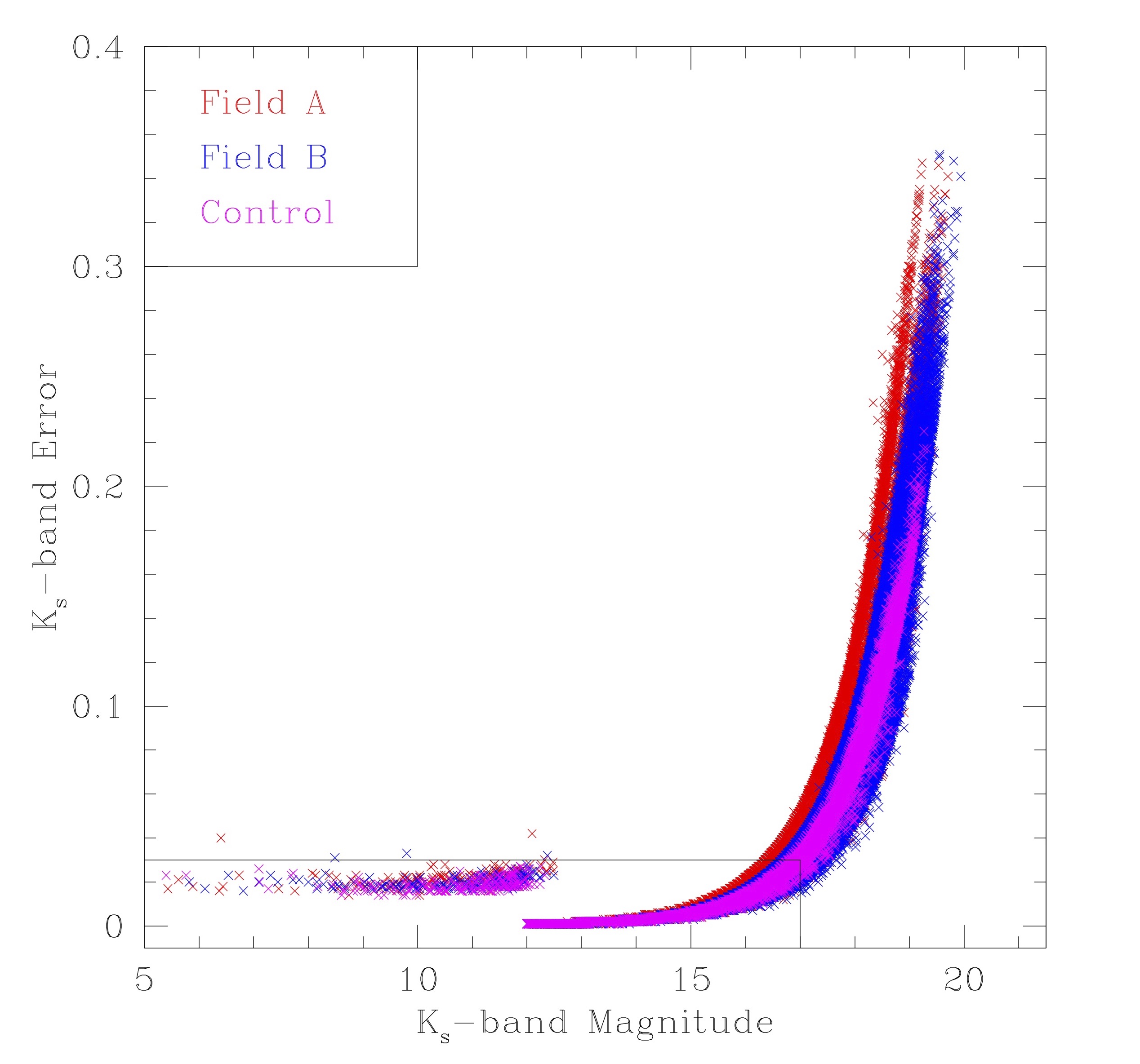}
\caption{\label{fig:errors} Error plots for the photometry.  The brighter stars come from the 2MASS survey \citep{2Mass}; at a given magnitude their errors will be higher than that of our deeper UKIRT data. The lines show our goal of a S/N of 30 (error 0.03~mag) at $J=18$ and $K=17$.}
\end{figure}

We show the final error plots in Figure~\ref{fig:errors}. We have drawn lines corresponding to our goal of a S/N of 30 (error 0.03~mag) at $J=18.0$ and $K_s=17.0$, and see that we have achieved this nicely. The 2MASS data at the bright end have larger errors, but this is somewhat deceptive as the uncertainties are not dominated by photon statistics. However, for our purposes these are moot as they are much brighter than any of the actual RSGs in our sample but are foreground objects.

\section{Identifying the Red Supergiants}
Our total sample consisted of 116,692 stars in Field A, 132,837 stars in Field B, and 11,254 stars in the control field,  for a total of 260,783 stars with $J$ and $K_s$ photometry.   (We provide the full photometry lists in Tables~\ref{tab:PhotAB} and \ref{tab:PhotC}.) We expect the RSG population to be a tiny fraction of this sample.  This section describes how we went about identifying it.

In Fig.~\ref{fig:CMD} we show the color-magnitude diagrams (CMDs) for each of the three fields. RSGs typically have effective temperatures of 3500-4100~K, corresponding to unreddened colors $0.85<(J-K)_0<$1.20 at M31's metallicity  \citep{Levesque2006}.  \cite{MasseySilva} has measured the temperatures of 16 M31 RSGs using spectral fitting of the TiO bands employing a new generation of the MARCS stellar atmospheres \citep{Marcs75,Marcs92}, finding typical $A_V\sim$1~mag, corresponding to an $E(B-V)\sim0.3$ or $E(J-K)\sim0.2$.  (We will delve more deeply into the reddening issue in Section~\ref{Sec-transforms}).  Thus, {\it roughly}, we would expect to find the RSGs between $J-K_s$ of 1.0 and 1.3 (warm/cool).  As argued above, we would like to extend our detection down to $K_s\sim16.2-16.7$ (cool/warm).  The most luminous RSGs have $\log L/L_\odot$ of 5.5 (see Fig.~\ref{fig:Evans} here and Fig.\~3 in \citealt{Levesque2005} corresponding roughly to $K\sim12.5-13.0$ (cool/warm). We have outlined the relevant region in the CMDs in red.  We see that the actual sequence of RSGs is a bit tilted with respect to this parallelogram, but it adequately guides the eye to the correct section of the CMD.

There will be two sources of contamination that we must concern ourselves with. The first of these is that of foreground stars; the second is that due to M31's own asymptotic giant branch (AGB) stars.    We deal with both of these in the next two sections.

\subsection{Cleaning the Sample by Removing the Foreground Stars}
\label{Sec-gaia}
The first contaminant to deal with is foreground stars: red dwarfs in the Milky Way's disk, and red giants in the Milky Way's halo, can have magnitudes and colors like those of M31's RSGs (see discussion in \citealt{MasseySilva}).  {\it Gaia} astrometry---both proper motions and parallaxes---provides us the tools for removing the first of these; the second of these is a negligible constituent as we show below.  

\begin{figure}
\epsscale{0.48}
\plotone{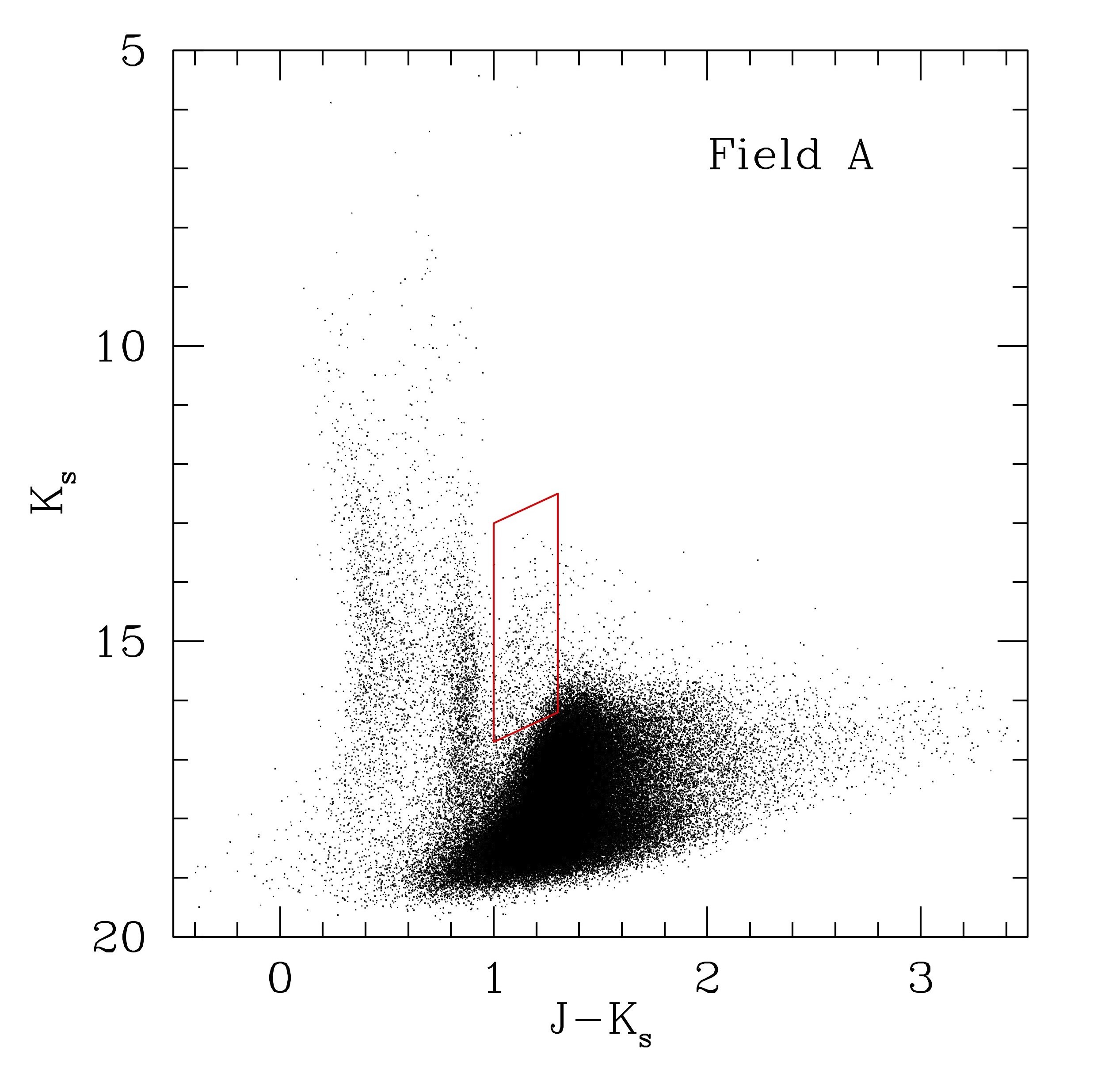}
\plotone{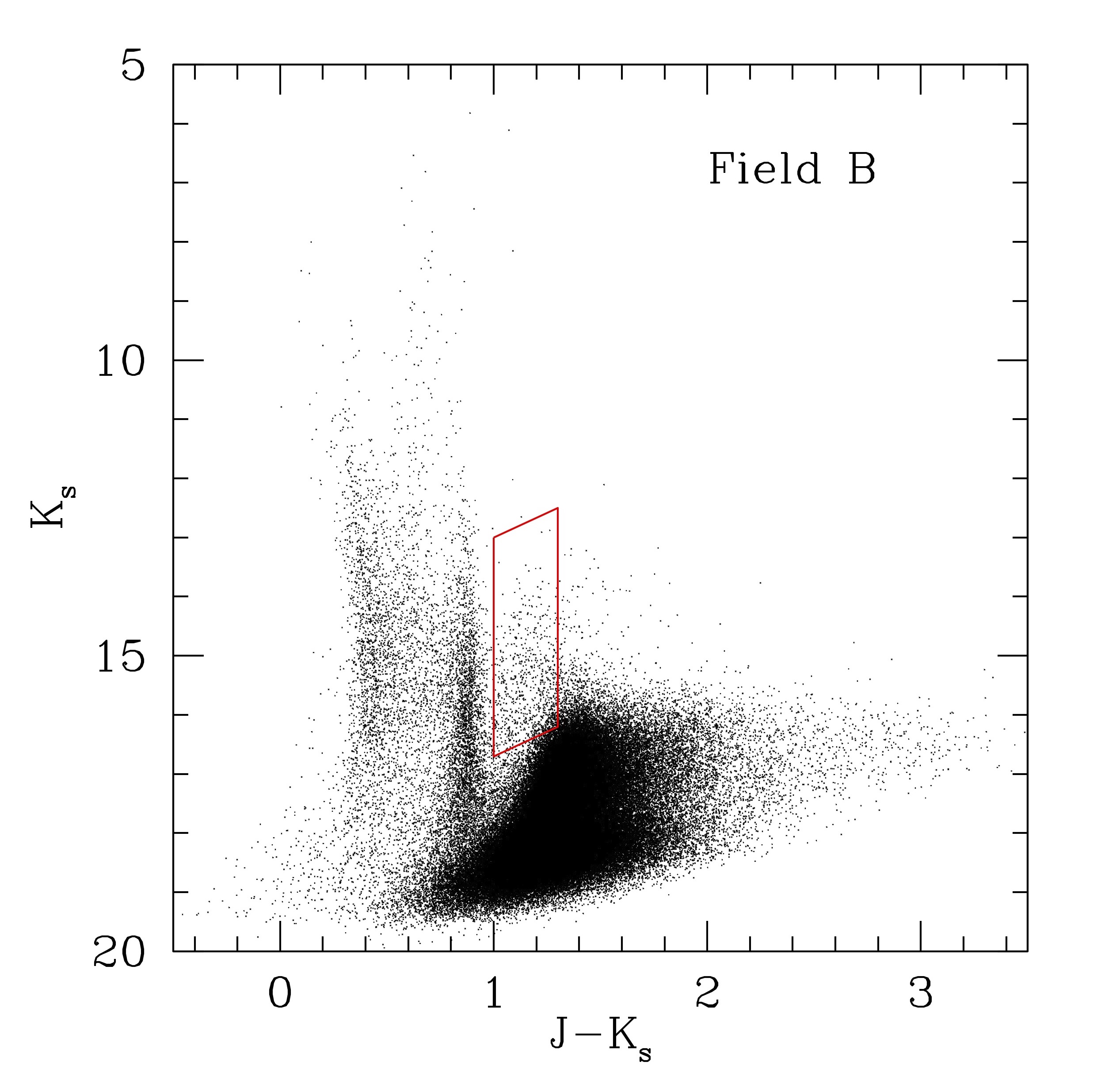}
\plotone{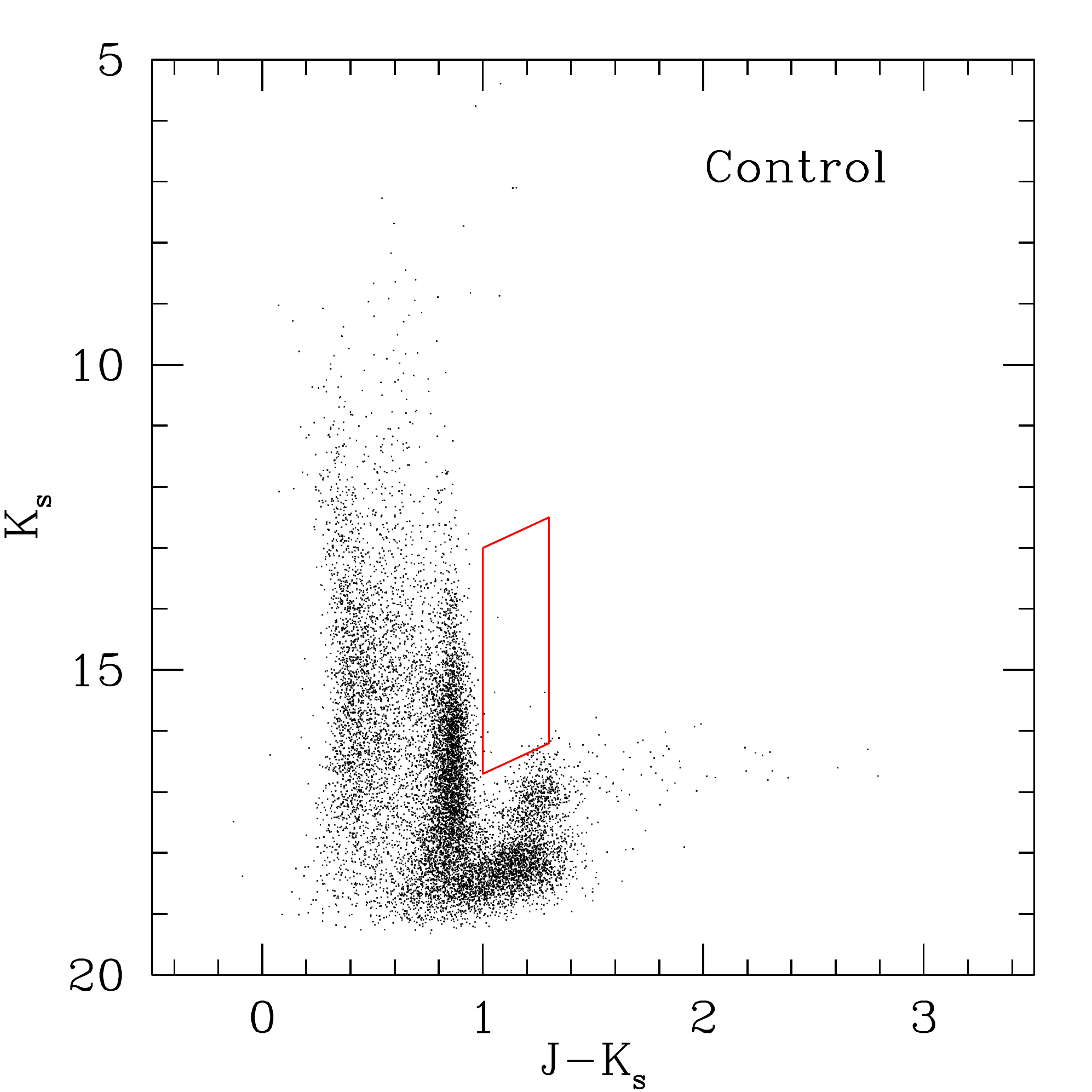}
\caption{\label{fig:CMD} The observed CMD for all three fields.  The UKIRT photometry, supplemented by 2MASS values at $K_s<12$ and $J<12$ as explained in the text.   The control number and location of stars in the control field gives a good impression of the degree of foreground contamination in the two M31 fields; recall that the area of Field A is only 75\% of that of Field B and the control field.  The red parallelogram shows the expected region of RSGs, based upon temperatures of 3500-4100~K, reddenings corresponding to $A_V=1.0$mag, and $4.2<\log L/L_\odot < 5.5$.
}
\end{figure}

We can gather a quick visual impression of where foreground contamination is significant by comparing the number and location of stars in the control field to that of the two M31 fields in Figure~\ref{fig:CMD}.   Few, if any, foreground stars are to be found in the region of the RSGs, while slightly warmer stars will be dominated by foreground stars.  This warmer sequence is where the yellow supergiants (YSGs) are found, and as we have previously shown by radial velocity studies, the YSGs are a rare constituent ($\sim$5\%) in that part of the CMD \citep{Drout2009}.  Stars of similar colors to our RSGs but fainter are also expected to have a strong foreground component, as seen in Figure~\ref{fig:CMD}.   

\subsubsection{Defining the {\it Gaia} Membership Criteria}
We explicitly remove foreground contamination in by utilizing the proper motions ($\mu_{\alpha}$, $\mu_{\delta}$) and parallaxes ($\pi$) from the \emph{Gaia} Data Release 2 (DR2). In order to astrometrically distinguish probable M31 members from likely foreground stars we follow a procedure based on that described in \citet{Gaia2018} for the LMC/SMC. In summary: we first define a three dimensional filter in $\mu_{\alpha}$, $\mu_{\delta}$, and $\pi$ based on a sample highly probable members of M31, and this filter is subsequently applied to the full set of UKIRT sources described above in order to assess their consistency with the expected proper motions and parallaxes of M31 members.

In order to define our filter, we begin by selecting a sample of stars from the Local Group Galaxy Survey of M31 \citep{LGGSII}. We restrict ourselves to stars with B-V $>$ 1.8 and B-V $<$ 0.3 in order to eliminate the yellow region of the color magnitude diagram, which has been shown to be $>$95\% contaminated by foreground dwarfs \cite{Drout2009}. These stars are then cross-matched with \emph{Gaia} DR2, and we further restrict the sample to stars with G$<$19.5 and $\pi$/$\sigma_{\pi}$ $<$5. We determine the median proper motions and median parallax for this sample and, finally, following \citet{Gaia2018}, we further eliminate any sources whose $\mu_{\alpha}$ or $\mu_{\delta}$ are more than four times the robust scatter estimate in order to minimize any remaining contributions from foreground stars.  

Applying these cuts, we are left with a sample of $\sim$2200 stars, which we use to determine the covariance matrix {\bf $\mu$} of $\mu_{\alpha}$, $\mu_{\delta}$, and $\pi$. Based on this matrix, we define multiple filters that will be used to classify our UKIRT sources as either probable M31 members or likely foreground dwarfs. Specifically, if:

\begin{itemize}
\item {\bf $\mu^{T} \sigma^{-1} \mu$} $>$ 12.8, a star is classified as a probable foreground star. This value corresponds to the 99.5\% confidence region. 
\item {\bf $\mu^{T} \sigma^{-1} \mu$} $<$ 4.11, a star is classified as a likely M31 member. This value corresponds to the 75\% confidence region.
\item 4.11 $<$ {\bf $\mu^{T} \sigma^{-1} \mu$} $<$ 12.8, a star is classified as having uncertain membership.
\end{itemize}

\subsubsection{Eliminating the Foreground Stars}
We apply these filters to our sample of UKIRT sources (in both the M31 and control fields), after cross-matching with the \emph{Gaia} DR2 database.   Of course, not all of our sources have measurements from {\it Gaia}---in particular, the fainter, reddest stars have no {\it Gaia} data. 

\begin{figure}
\epsscale{0.48}
\plotone{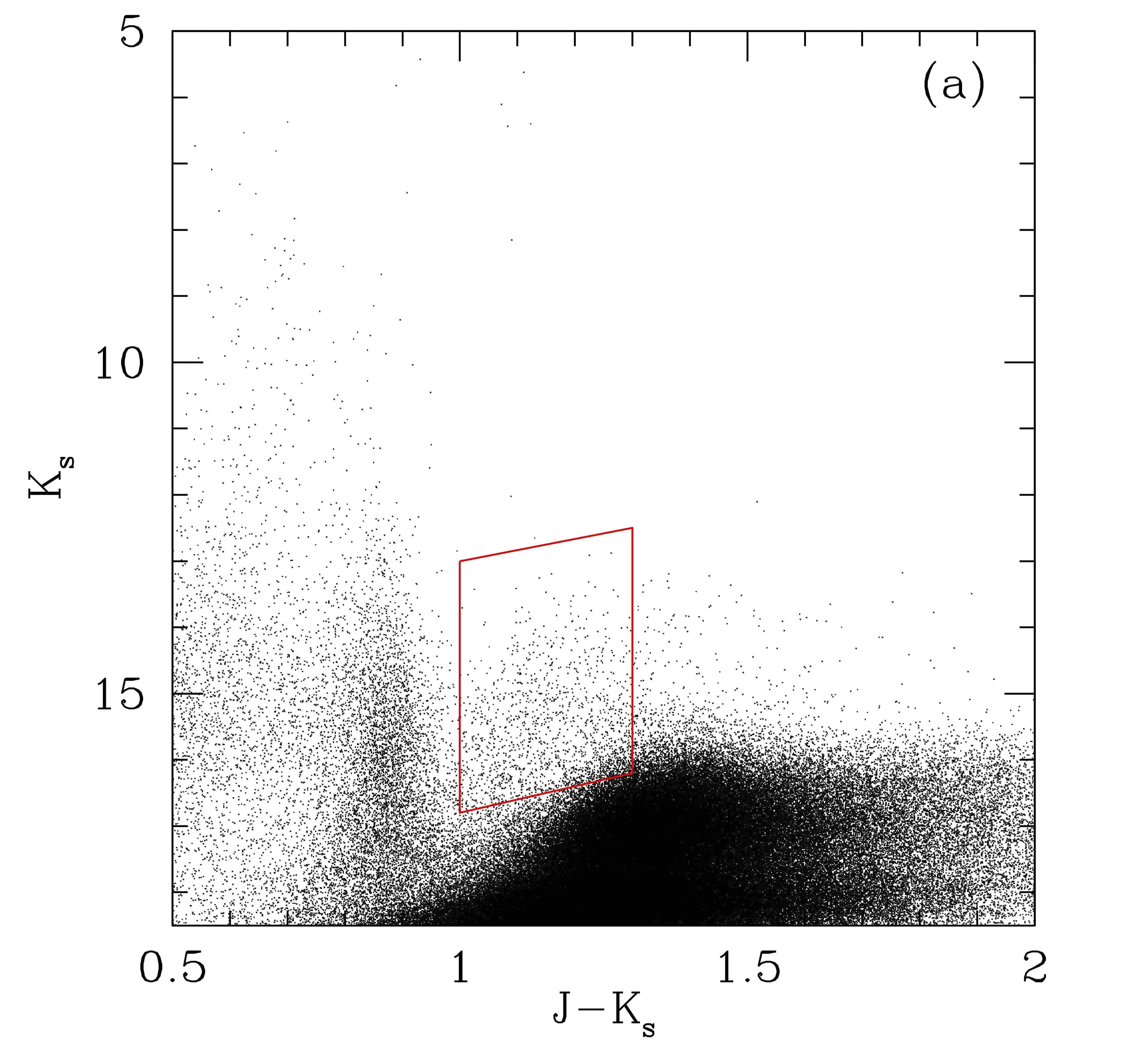}
\plotone{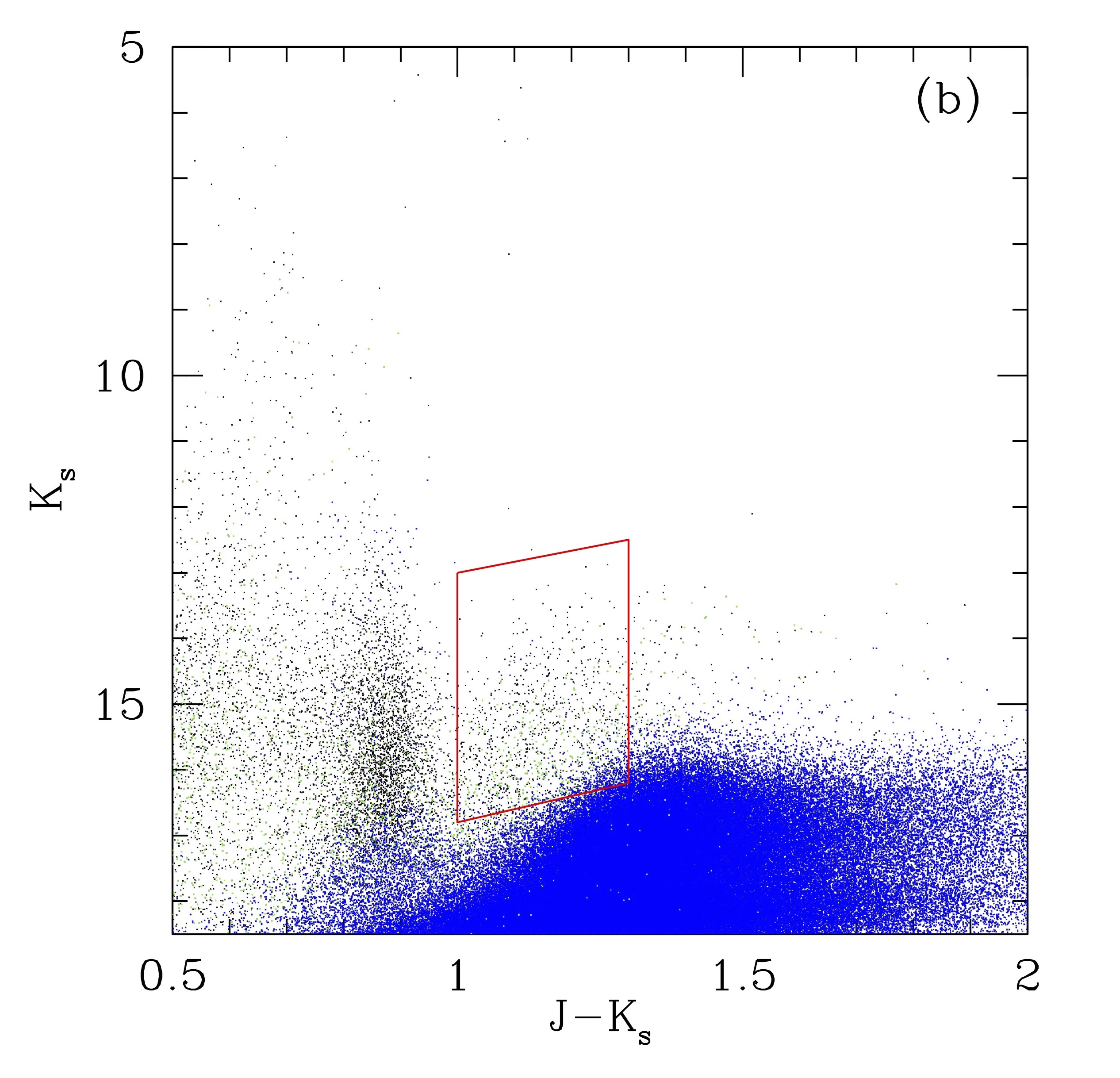}
\plotone{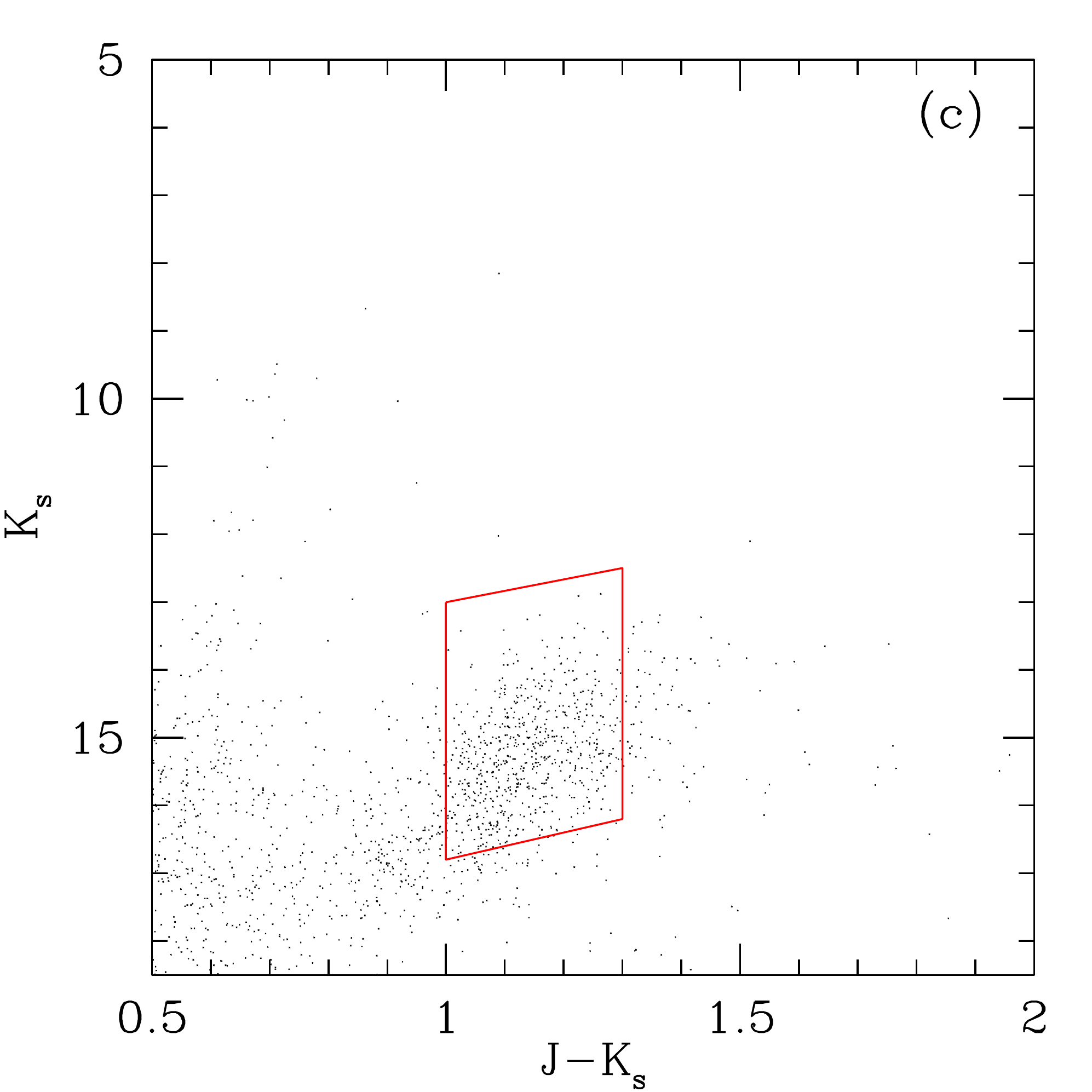}
\plotone{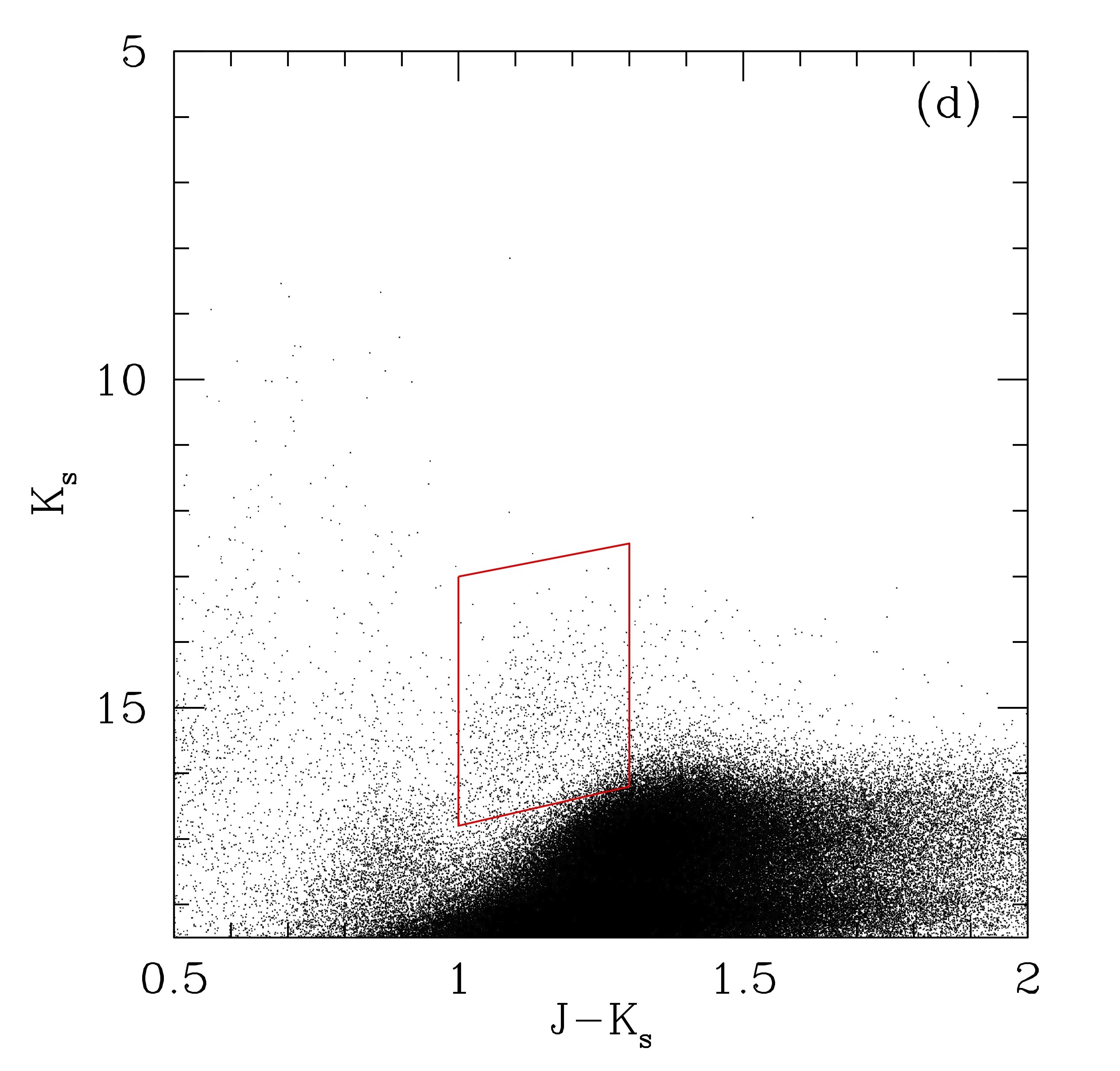}
\caption{\label{fig:CMDGaia}  The CMD for the two M31 fields combined.  In (a) we show all of the data, with the parallelogram showing the approximate location of where we expect RSGs to be found.  In (b) we color-code the data based upon whether or not the star's membership status can be determined from {\it Gaia} or not; the blue points have no {\it Gaia} data, and the green points show ambiguous membership.  In (c) we show the CMD for the confirmed members only.  In (d) we now only exclude the stars for which {\it Gaia} shows that they are clearly non-members.
}
\end{figure}

In Fig.~\ref{fig:CMDGaia}(a) we show a closeup of the CMD,  again indicating the general region where we expect RSGs to be found.  In Fig~\ref{fig:CMDGaia}(b) we show stars with no Gaia information (blue) and those with uncertain membership status based on Gaia data (green).  Finally in Fig.~\ref{fig:CMDGaia}(c) we show only the stars certain to be members, while in Fig.~\ref{fig:CMDGaia}(d) we show all of the data {\it except} the stars certain to be foreground. 

\begin{figure}
\epsscale{0.48}
\plotone{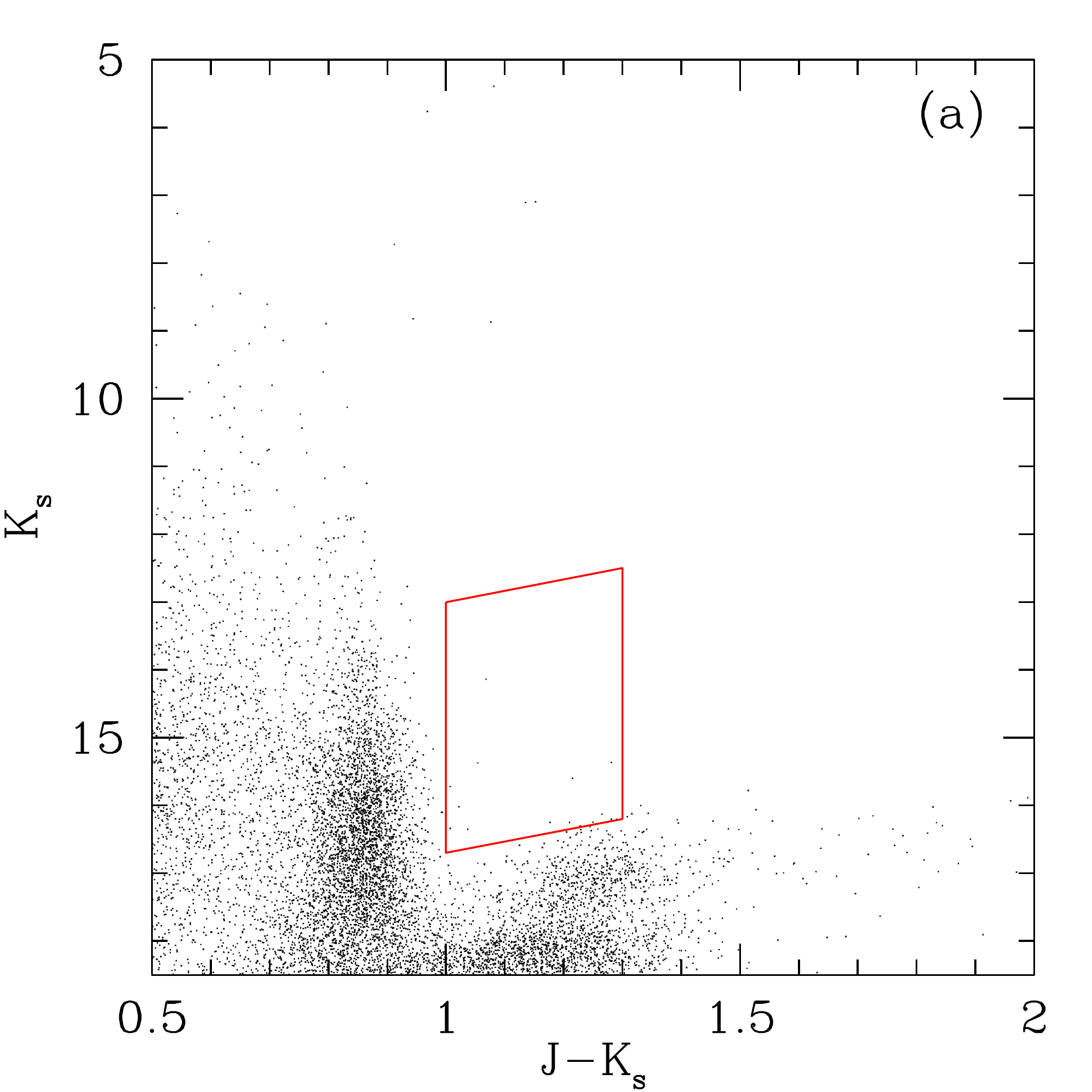}
\plotone{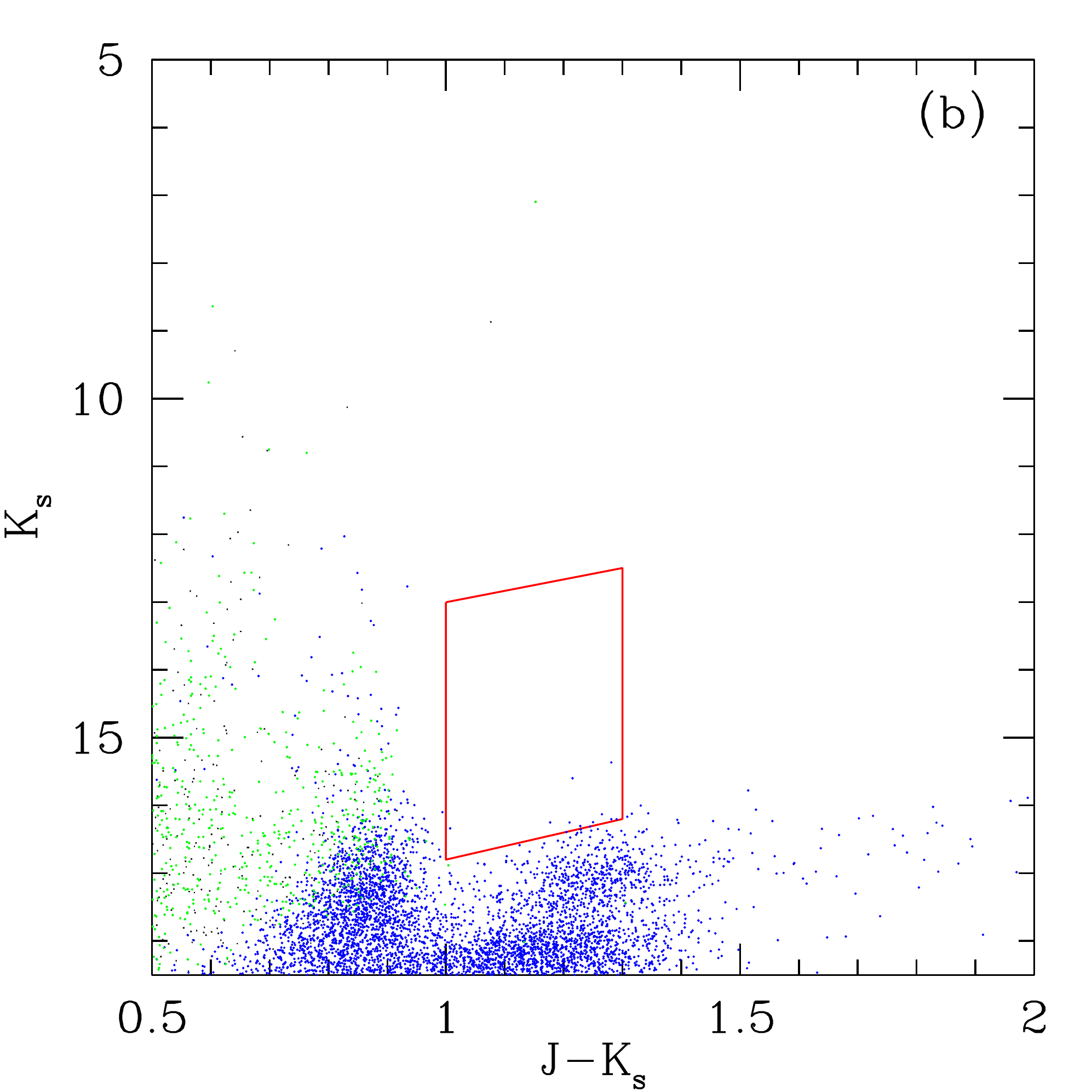}
\caption{\label{fig:ForegroundCleaned}  The CMD for the control field.  In (a) we show all of the data, with the parallelogram showing the approximate location of where we expect RSGs to be found.  In (b) we have removed the {\it Gaia}-selected foreground stars, and color-coded the data based upon whether or not the star's membership status can be determined.  As with the previous figures, the blue points have no {\it Gaia} data, and the green points show ambiguous membership.  
}
\end{figure}

What fraction of the stars with no {\it Gaia} data likely to be foreground?   We can answer this by appealing to the control field CMD. In Figure~\ref{fig:ForegroundCleaned} we show the control field before and after removing {\it Gaia}-identified foreground stars.  We have once again indicated the stars with no {\it Gaia} information by color-coding the points blue, and the ones with ambiguous results, green.   By comparing Figure~\ref{fig:ForegroundCleaned} (a) with \ref{fig:ForegroundCleaned}(b), we see that most of the handful of stars in the RSG region on the control field (which should all be foreground stars) are correctly identified as foreground stars based on their kinematics. The remaining ones have no {\it Gaia} data.  At fainter magnitudes ($K_s<16$) in the same color range there are also no {\it Gaia} data.  We will revisit this issue once we better define the region of RSGs.
 
We note that the control field also has proven useful in confirming that the number of halo giants contaminating our data is negligible.  Giants in the MW's halo, at distances of 10~kpc or more, will have parallaxes and proper motions so tiny that even {\it Gaia} data (at present) has a hard time distinguishing them from members of other galaxies, at least on a star-by-star basis\footnote{Very careful analysis has allowed the tangental rotation of M31 and M33 to be measured in a {\it statistical} sense;  see \citet{MarelM31}.}  The lack of any stars identified as members or with ambiguous information in the RSG region of the control field CMD shows that this is not a problem. 

\subsection{Removing the AGB Stars}
\label{Sec-AGBs}

The other contaminant is M31's own asymptotic giant branch (AGB) stars.  These are the the bane of RSG population studies at lower luminosities.  AGB stars are evolved low- to intermediate-mass stars that are past their core-He burning phase but undergoing He- and H-shell burning.  Their temperatures are generally cooler than that of RSGs, and their luminosities overlap those of the lower mass RSGs. \citet{Brunish86} may have been the first to emphasize the potential confusion between intermediate-mass AGB stars and RSGs in stellar population studies, and suggested that at luminosities below $\log L/L_\odot=4.9$ there could be overlap in luminosity. Studies like \citet{MasseyOlsen} often adopt this as a cut-off; that luminosity roughly corresponds to 20$M_\odot$ (see, e.g., Figure 2 in \citealt{Ekstrom2012}). Of course, we would like to go much lower than that.

However, with good NIR colors it is relatively easy to separate RSGs and AGBs at lower luminosities, as AGBs are significantly cooler.   \citet{Cioni2006a,Cioni2006b} and \citet{Boyer2011} showed that one could in fact readily separate the various types of AGBs (carbon-rich, oxygen-rich) from one another, as well as separate the AGBs from RSGs.  This has most recently been used by  \cite{Yang2019} in the SMC, who claim unambiguous separation of RSGs from AGBs down to the tip of the red giant branch at about $K_s=12.7$ in the SMC, which would roughly correspond to $K_s=18.1$ in M31.  This is 1.5~mag deeper than we need to go to achieve $\log L/L_\odot=4.0$.

\begin{figure}
\epsscale{0.45}
\plotone{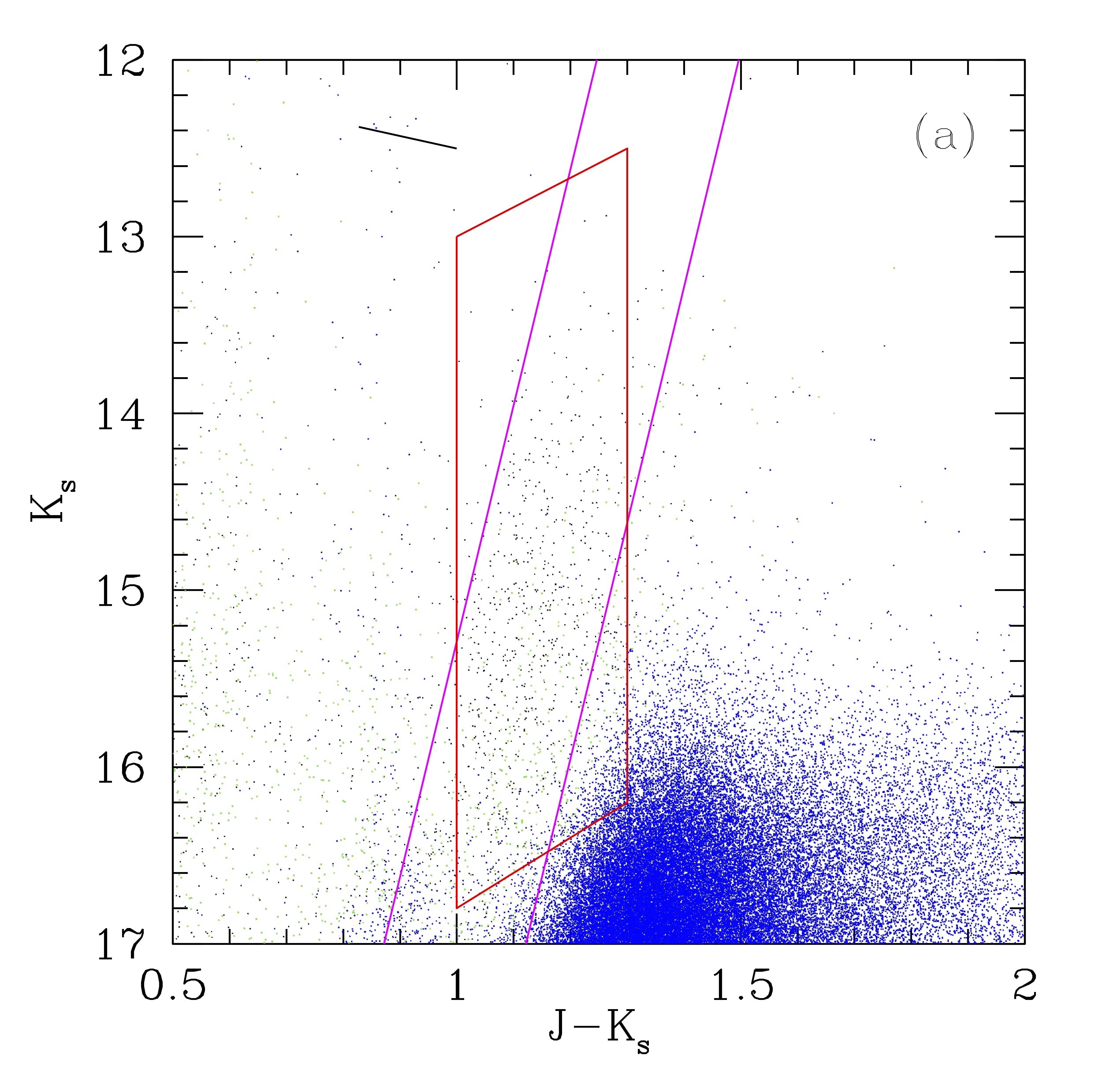}
\plotone{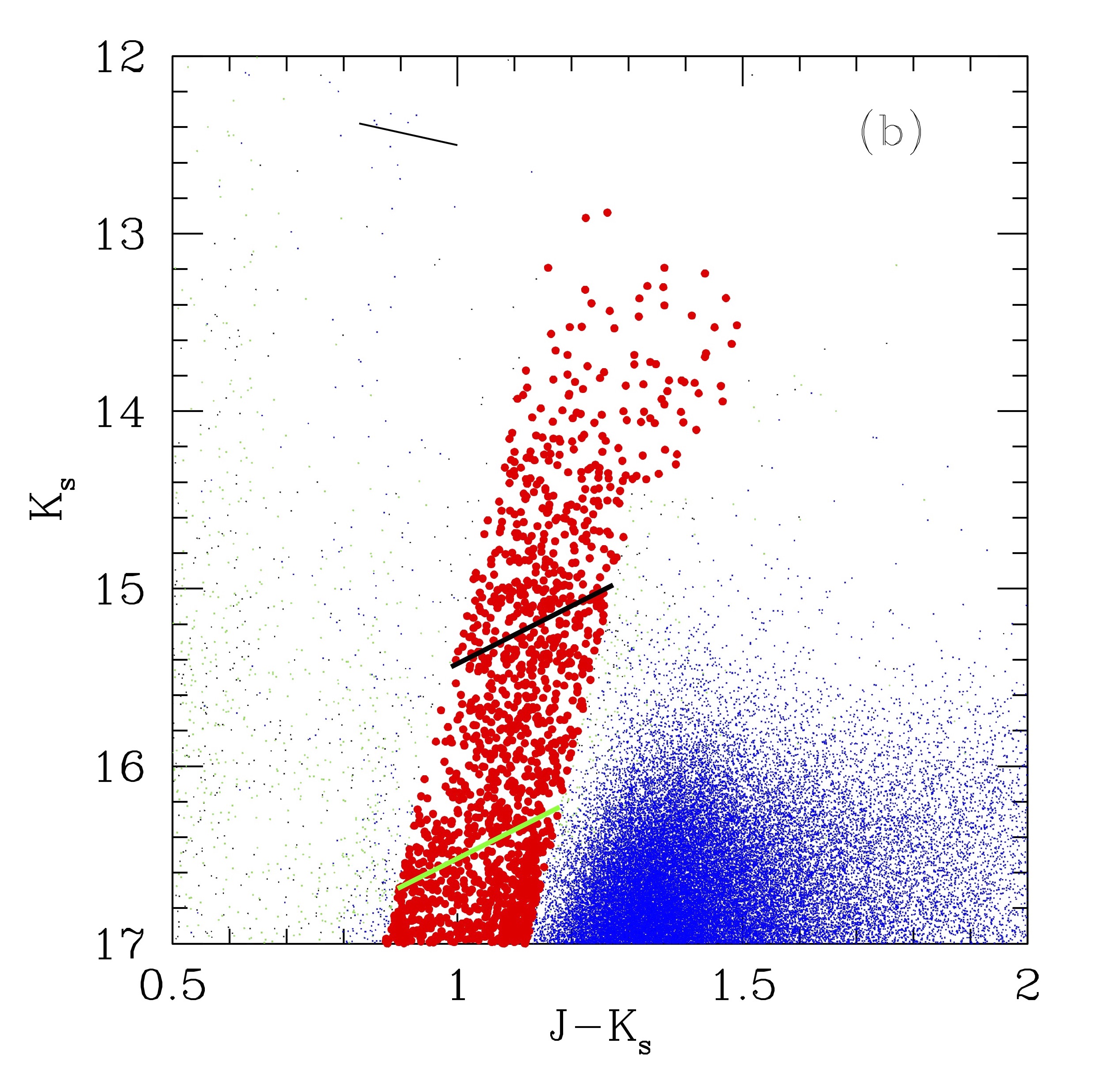}
\plotone{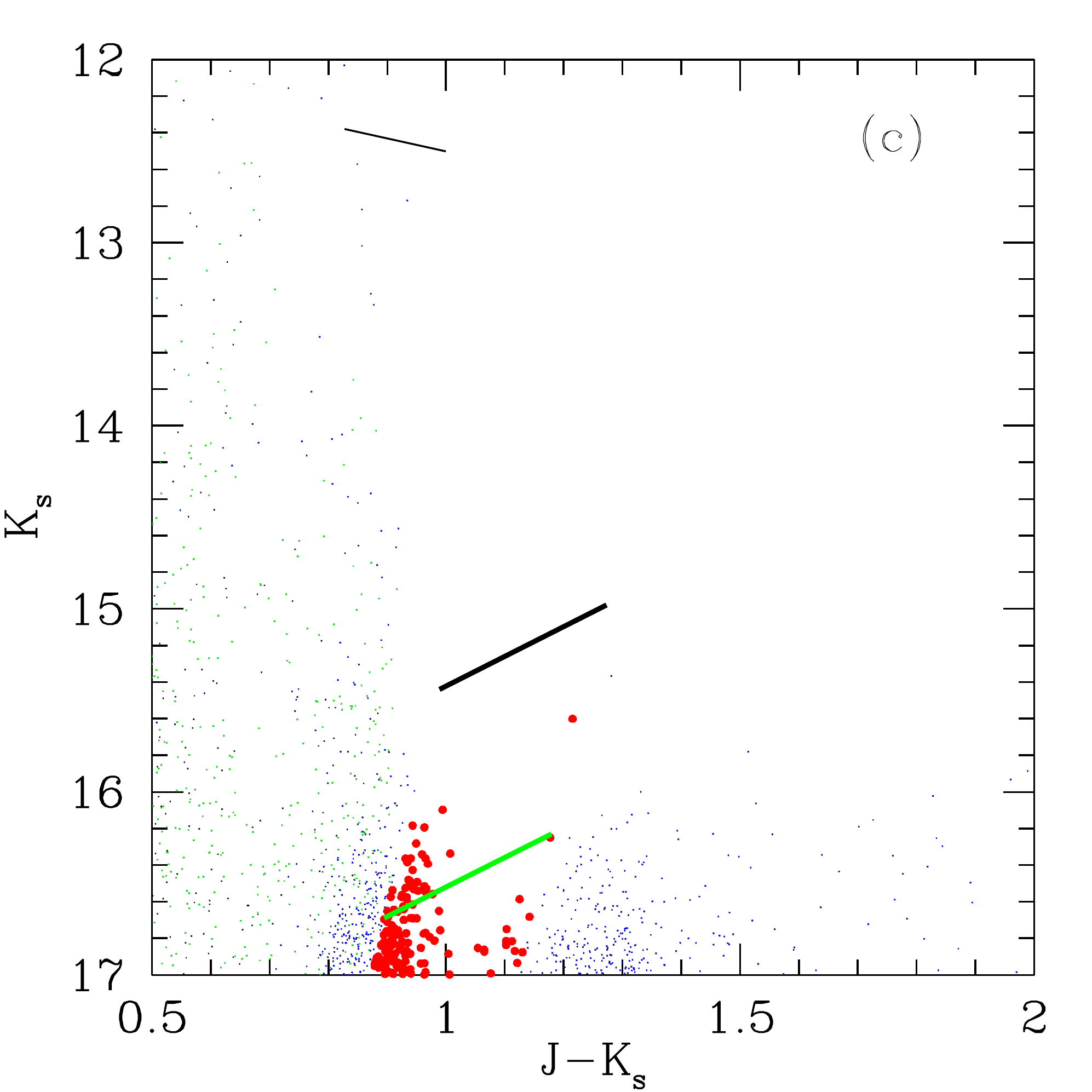}
\caption{\label{fig:Ming} The Identification of RSGs.  (a) The two magenta lines denote the region expected to be RSGs based on the ``CB" method of \cite{Yang2019}, with a 0.16~mag adjustment in $(J-K_s)$ due to the expected shift in the temperature of RSGs from the SMC to M31, and a 5.55~mag adjustment in $K_s$ due to the greater distance of M31 compared to the SMC.  The blue points are stars without membership data, and the green points are the ones with ambiguous membership data.  Non-members have been removed.  The line at upper left shows the reddening vector corresponding to $A_V=1.0$ mag.   (b) The same as (a), but now with the adopted RSGs indicated by larger red points.  The thick green diagonal line going through the RSG points denotes $\log L/L_\odot=4.0$, our limit.  For comparison $\log L/L_\odot=4.5$ is denoted by the black line.   (c) The same as (b) but for the control field.  The ``RSGs" in this field (large red points) are likely all foreground objects, and their numbers will be used to correct the luminosity function we derive.
}
\end{figure}

In Fig.~\ref{fig:Ming} (a) the two diagonal magenta lines show the \cite{Yang2019} color cuts for RSGs using their ``CB" method (i.e., based on \citealt{Cioni2006a,Cioni2006b} and \citealt{Boyer2011}), adjusted for the difference in distances (59~kpc for the SMC and 760~kpc for M31, according to \citealt{vandenbergh2000} and references therein) and by 0.16~mag in $J-K_s$.  The latter was determined first empirically by eye, but is consistent with what we expect due to the change of the Hayashi limit with metallicity.  The typical RSG in the SMC is a K2-3~I with a typical temperature of 4000~K (\citealt{Levesque2006}, Table 4).  According to the MARCS models, this corresponds to a $(J-K)_0$ of $0.951$ at SMC metallicity.  The typical RSG in M31 is an M2~I with a temperature of 3600~K (\citealt{MasseySilva}, Table 5).  According to the MARCS models this correspond to a $(J-K)_0$ of 1.121.  Converting this 0.17 difference in $(J-K)_0$ translates to a difference of $0.16$ in $ (J-K_s)_0$ according to the transformation equation of \cite{Carpenter}, as discussed below in Sec.~\ref{Sec-transforms}, a rather remarkable, and perhaps fortuitous, agreement.  We defined the width of the RSG band to be 0.25~mag wide in $J-K_s$, slightly larger than the 0.20~wide boundary adopted by \citealt{Boyer2011}. We list the equations for these two lines in Table~\ref{tab:FunFacts}. The blue points in that figure show the stars for which there are no {\it Gaia} data; this set is dominate by the AGBs. We see that that the vast majority of points to the left of our parallelogram have ambiguous membership; we believe most of these stars are foreground.  

Based on this we now identify the RSGs in Fig.~\ref{fig:Ming} (b).  We have relaxed the upper temperature limit slightly, and at brighter magnitudes we accept slightly cooler stars.  As \cite{Yang2019} argues this could contaminate the upper luminosity sample lightly by super-AGBs but the degree should be small and allows for the existence of dusty high-luminosity RSGs.     The adopted limits are as follows:
\begin{itemize}
\item $0.87 \leq (J-K_s) \leq 1.15$, $K_s\leq 17.0$ and $K_s < K_1$
\item $1.15 < (J-K_s) \le 1.32$, $K_s \leq 31.873-13.235 (J-K_s)$
\item $1.32 < (J-K_s) \leq 1.50$, $K_s \leq 14.40$
\end{itemize}
We include members, uncertain members, and stars without {\it Gaia} information, excluding only the {\it Gaia}-identified foreground stars.  The green line corresponds to our luminosity cutoff $\log L/L_\odot >4.0$; we can see that there is still a considerable gap between the coolest star and the mountain of AGBs.

In Fig.~\ref{fig:Ming}(c) we show the same information for the control field. We see that within the RSG region there is considerable contamination at the fainter magnitudes and warmer temperatures.  (This contamination is less than a few percent for $K_s<16$ but 30\% at the faintest magnitudes we include in determining our M31 RSG luminosity function.) We could simply correct our luminosity function for these stars, under the assumption that they are all foreground. However,  an examination of the spatial distribution of these control field stars show that there are $2.3\times$ more control field stars in the RSG region with $\log L/L_\odot \geq 4.0$ than on the side away from it, using the conversions described below.  We will count only the stars on the western half of the control field in correcting our RSG luminosity function (and of course scaling their numbers by the relative areas), and keep in mind that even so we may be slightly overestimating the size of the correction. 

\section{Transformations of Photometry to Effective Temperatures and Bolometric Luminosities}
\label{Sec-transforms}
In order to construct a luminosity function, we must use our photometry to determine bolometric luminosities; this requires that we derive effective temperatures and the resulting bolometric corrections.  Our first step is to transform our UKIRT photometry (tied to the 2MASS J, $K_s$ system) to the standard \citet{BessellBrett} (BB) system, as these standard bandpasses were used in computing the expected colors from the MARCS stellar atmospheres (see \citealt{Bessell}). \citet{Carpenter} provides transformations from standard system to 2MASS. Inverting these we find: $$(J-K)_{\rm BB}=\frac{[(J-K_{\rm 2MASS}+0.11]}{0.972}$$ and $$K_{\rm BB}=K_{\rm 2MASS}-0.044.$$

The next step was to correct the photometry for reddening. \citet{LGGSII} finds that the average reddening for OB stars in M31 is $E(B-V)=0.13$, or $A_V=0.40$.  However, as mentioned above, RSGs are usually considerably more reddened than their neighboring OB stars due to circumstellar dust \citep{Smoke}.  We would ideally like to correct the photometry for each star individually.  We have succeeded in that by fitting the spectra with MARCs models (i.e., \citealt{Levesque2005,Levesque2006,MasseySilva}).  We could simply adopt a single ``moderate" value for the extinction, say, $A_V=1.0$, but this overlooks the fact that in general the higher luminosity RSGs are more heavily reddened than lower luminosity RSGs. For instance,  the M31 stars analyzed by \cite{MasseySilva} show a strong trend in $A_V$ with bolometric luminosity, with the more luminous stars showing the greater reddening.  

\begin{figure}
\epsscale{0.45}
\plotone{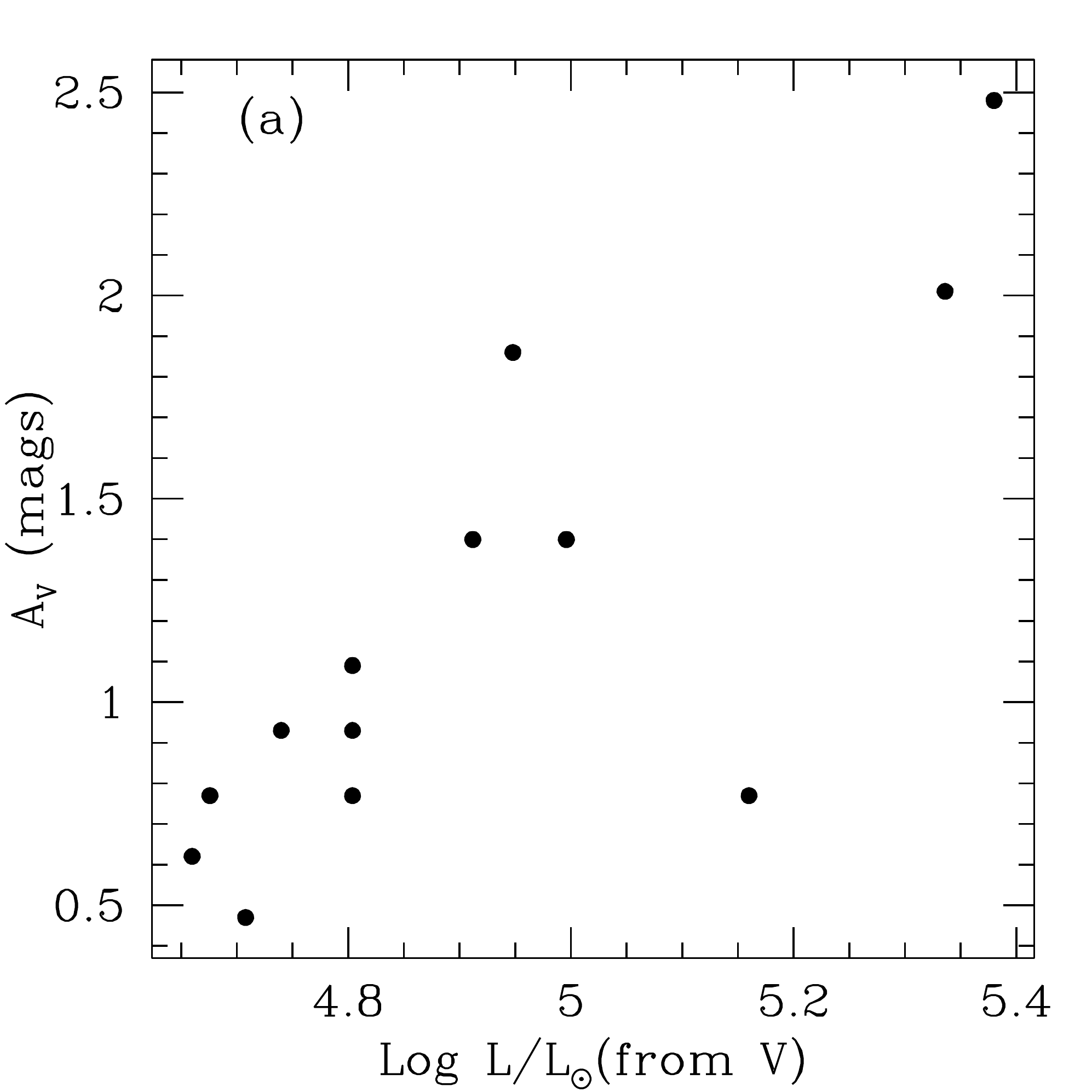}
\plotone{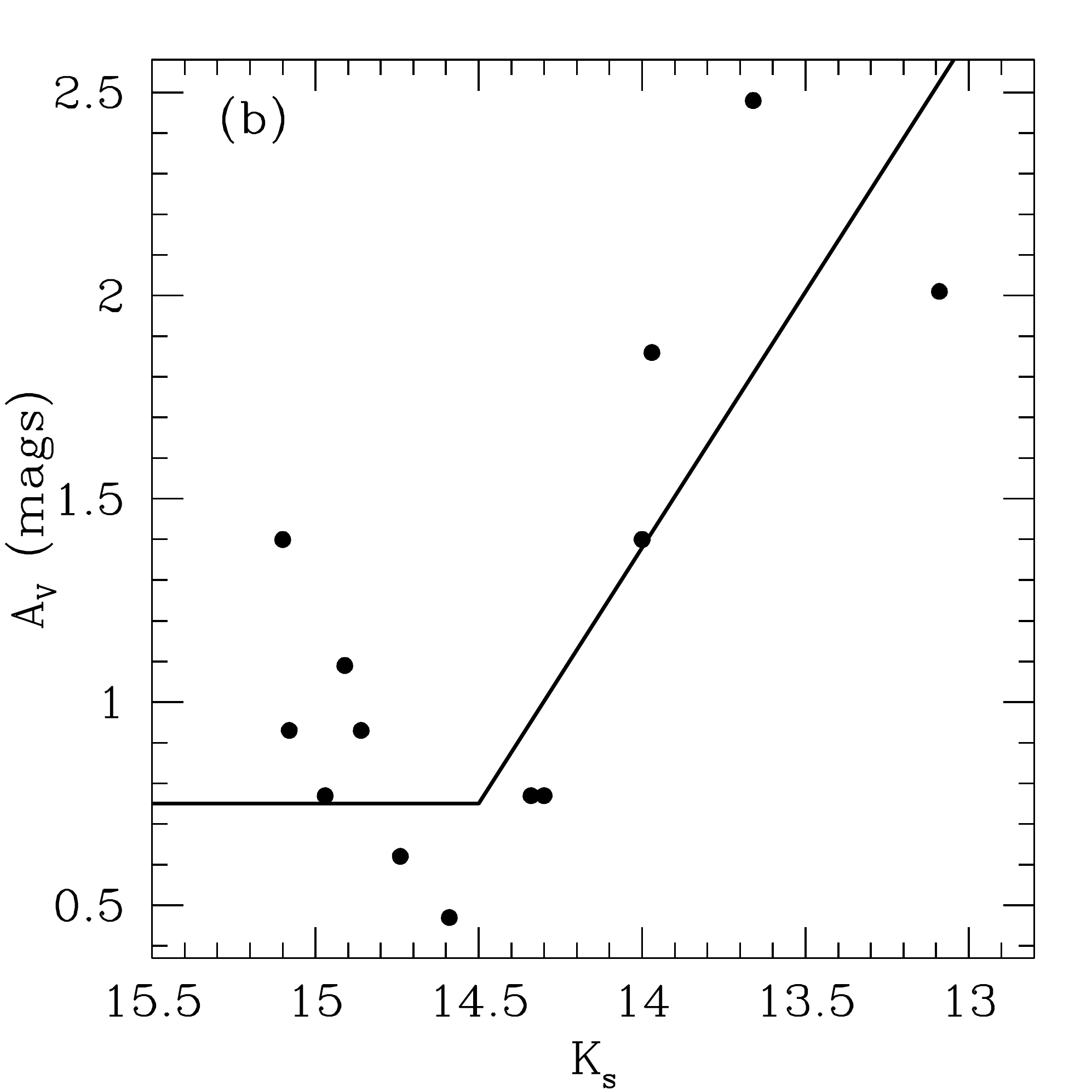}
\caption{\label{fig:Av} 
Extinction of Previously Analyzed M31 RSGs. (a) The extinction $A_V$ is shown as a function of the bolometric luminosity for the sample of stars analyzed by \citet{MasseySilva}.  (b) Same as (a), but now with the extinction plotted as a function of $K_s$ magnitude as given by \citet{MasseySilva}.  The adopted extinction relation is shown by the solid line. }
\end{figure}

We illustrate the correlation in Fig.~\ref{fig:Av}(a).  To show this is not simply a logical tautology (since the stars' luminosities are computed using the adopted extinctions), we show in Fig~\ref{fig:Av}(b) a similar plot where we have replaced the abscissa values with the uncorrected $K_s$-band photometry.  Although the scatter is larger, particularly at the low luminosity end, the same trend is seen; RSGs in M31 that have brighter $K$ values show more extinction. Recall  that $A_K=0.12 A_V$ and $E(J-K)=A_V/5.8$; i.e., we expect a color excess of 0.10-0.20 in $J-K_s$ at low luminosities ($\log L/L_\odot<4.8$) but $0.35-0.45$ at higher luminosities.   

We have therefore decided to adopt a color-dependent relation for the extinction.  For $K_s \geq 14.5$ we adopt $A_V=0.75$ mag, i.e., $A_K=0.09$~mag, and $E(J-K)=0.13$.  For brighter stars, we adopt $A_V=0.75-1.26(K_s-14.5)$. For our brightest stars this translates to $A_V\sim 2.0$, or $A_K=0.25$~mag, with $E(J-K)=0.34$.  

We then transformed these magnitudes and colors to effective temperatures and bolometric luminosities via the MARCS models. As part of our previous work, our collaborator Bertrand Plez (private communication 2005, 2009) provided us with tables of colors and bolometric corrections for each model. We used these to derive new relationships based upon the solar metallicity models used by \citet{Levesque2005} and 1.8$\times$ solar models used by \citet{MasseySilva}; the best determinations of the H\,{\sc ii} region abundances in M31 suggest a metallicity of $\sim$1.5 solar based upon the oxygen abundances \citep{Sanders}. We restrict ourselves to the $\log g$=0.0 models as these are the most appropriate for the RSGs with known physical properties \citep{Levesque2005,Levesque2006,MasseySilva}.  We find that the relationships are quite linear over the relevant temperature range (3500-4500~K): $$T_{\rm eff}=5643.5 - 1807.1\times (J-K)_0$$ and $$BC_K=5.567 - 0.757 \times \frac{T_{\rm eff}}{1000}.$$

Our adopted distance to M31 of 760 kpc is equivalent to a true distance modulus of 24.40. Thus $$M_{\rm bol}=K_0+BC_K-24.40,$$ and $$\log L/L_\odot=\frac{M_{\rm bol}-4.75}{-2.5},$$ where we have adopted the bolometric luminosity of the sun as 4.75.

With these conversions, and our adopted reddenings, we now show in our CMD the limits corresponding to $\log L/L_\odot$ 3.95 (the lowest part of 0.1~dex wide bin centered on $\log L/L_\odot$=4.0). Recall from our discussion of our error budget in Section~\ref{Sec-obs}: our errors even at the faintest magnitude levels correspond to $<$100~K (about 0.01~dex) in temperature, and 0.04~dex in $\log L$.  

\section{Model Comparisons}
\begin{figure}
\epsscale{0.45}
\plotone{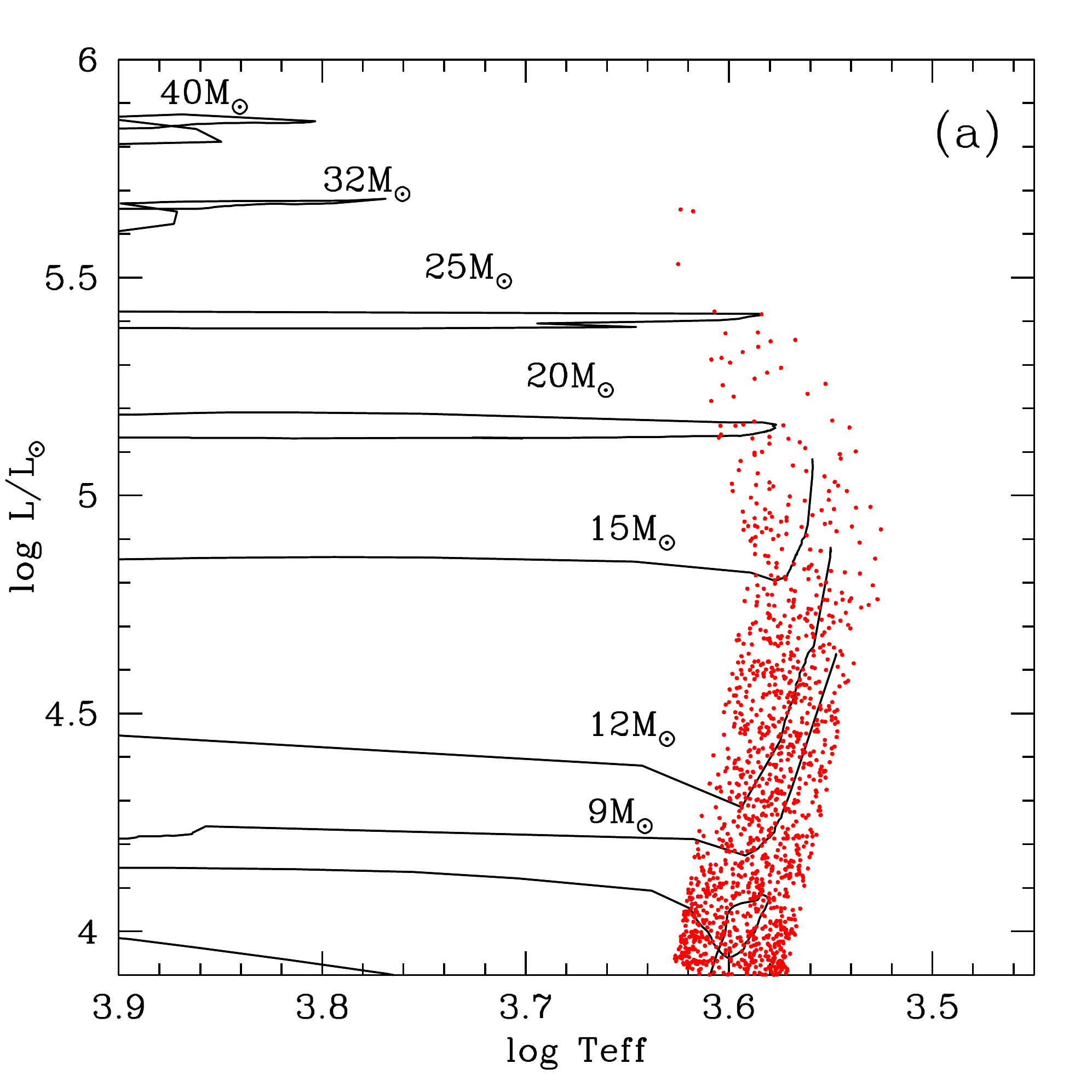}
\plotone{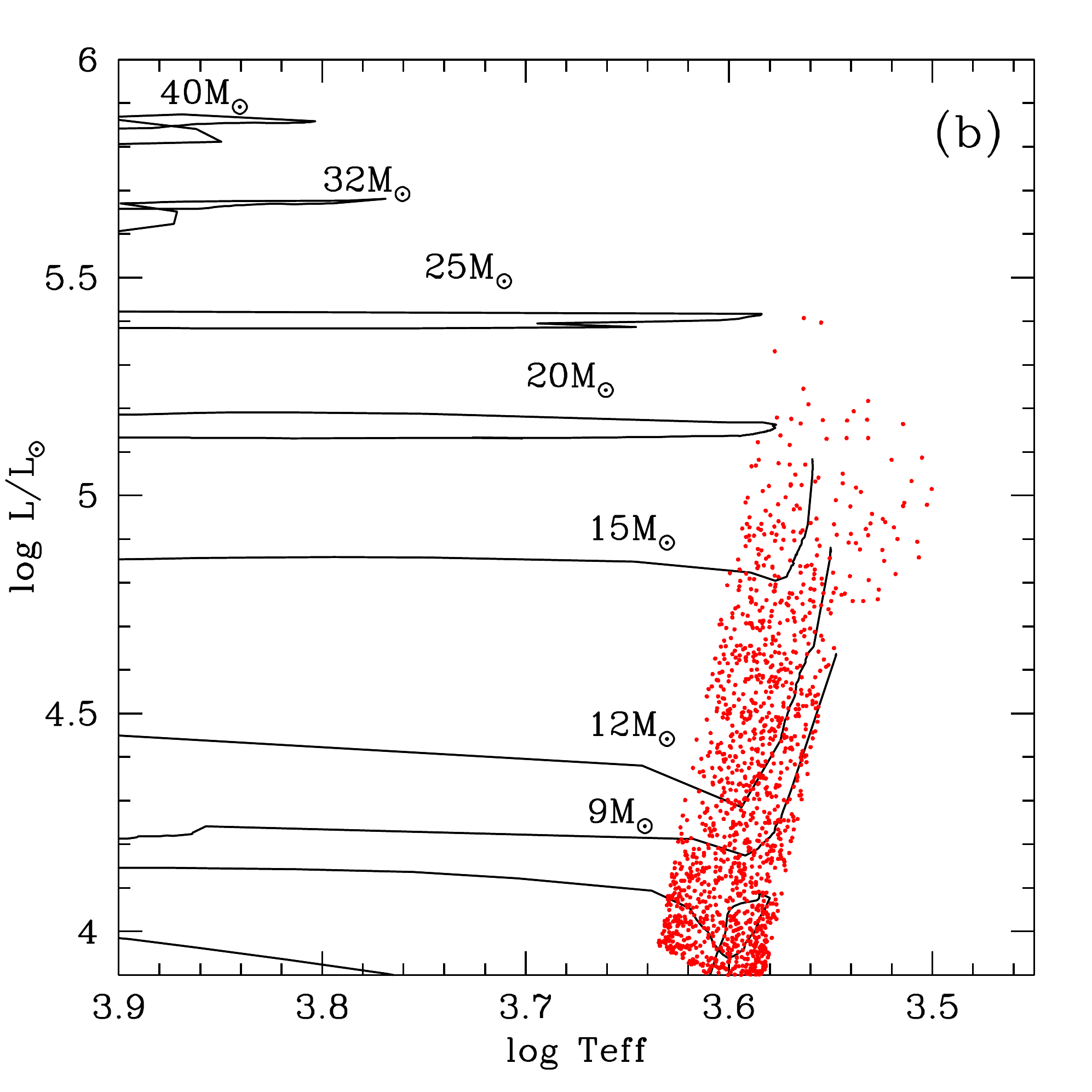}
\caption{\label{fig:HRD}
HRD for our RSG sample.  The tracks shown are from \citet{Ekstrom2012} computed for solar ($z=0.014$) metallicity and with an initial rotation of 40\% of the breakup speed at ZAMS.  (a) The location of our RSGs are shown using the adopted K-dependent relationship for the extinction. The turn towards higher temperatures at higher luminosities is consistent with the location of the tracks.  (b) The location of the RSGs assuming a constant value for the extinction (equivalent to $A_V=1.0$), similar to what was done by \cite{Massey2016}.  Note the deviation from the location of the tracks.} 
\end{figure}

Figure~\ref{fig:HRD} now shows the location of our RSG sample in the H-R diagram.   In (a) we have adopted the above relationship between the extinction and brightness.  We see that the location of the points are very good match to the location of the Geneva evolutionary tracks, and even turn back towards higher temperatures at higher luminosities. In (b) we adopt a constant value $A_V=1.0$~mag following \cite{Massey2016}.  The lower luminosity points are too warm compared to the tracks, and the higher luminosity points scatter to much cooler temperatures than the tracks.  

In (a) we do find three rather high luminosity RSGs with $\log L/L_{\odot} > 5.5$.  The {\it Gaia} data for all three of these strongly indicate membership in M31.  It is possible of course that for these stars our method for computing the extinction has resulted in an overestimate for both the temperature and luminosity.   Adopting an average $A_V=1.0$~mag for these stars would result in moving these points down to where the top three points are in (b), a significant change. The luminosity of the most luminous RSG is of considerable interest (see, e.g., \citealt{MasseySilva}), and we plan to obtain spectrophotometry of these three stars and measure their reddening accurately using their SEDs. Until then, we will treat the luminosities of those three stars as uncertain. If these high luminosities are correct, this stands at odds with the result found by \citet{Davies2018} that the maximum luminosity of RSGs is in the region of $\log L/L_{\odot} = 5.5$. \cite{Massey2009} found, and \citet{Davies2018} confirmed, that there does not appear to be a metallicity dependence on the maximum luminosity, and thus the RSGs we have discovered here in M31 should not be any more luminous than those found by \citet{Davies2018} in the Magellanic Clouds.  If anything, we would expect them to be less luminous, given the shift of the atmospheric  Eddington limit to lower values at higher metallicity; see discussion in \citet{Lamers2017}.  We will revisit this issue in a future paper after we have investigated the RSG content of other galaxies in the Local Group.

With a better understanding of the RSG population of M31 in hand, we compared our results with that of the SYnthetic CLusters Isochrones \& Stellar Tracks ({\sc syclist}) models created by the Geneva Evolutionary group to check which mass-loss rate during the RSG phase best fits the observed luminosity function. A full description of {\sc syclist} can be found in \citet{SYCLIST}. As one of its many functions, it computes stellar populations as a function of time based on evolutionary tracks that are given as input. The model we used included 1000 stars within the cluster and a Salpeter IMF \citep{IMF}. Our minimum mass of $8 M_\odot$ and maximum mass of $30 M_\odot$ came from the expected mass range of RSGs. For our calculation we ignored the effects of binaries. We compute the positions of stars in the HRD for single-aged populations with ages between 0 and $1.6\times10^8$ years (corresponding to the total lifetime of the lowest initial mass star considered here) with time steps of 32,000 years (making 5000 steps). The program outputs, among other things, the mass-loss rate vs.\ luminosity for 1, $10\times$ and $25\times$ the mass-loss rates described in \citet{Meynet2002}, following the arguments of \citet{Georgy2012} and \citet{Meynet2015}. To get the luminosity distribution functions for the continuous star formation hypothesis, the models at each time step are summed up. 

\begin{figure}
\epsscale{0.45}
\plotone{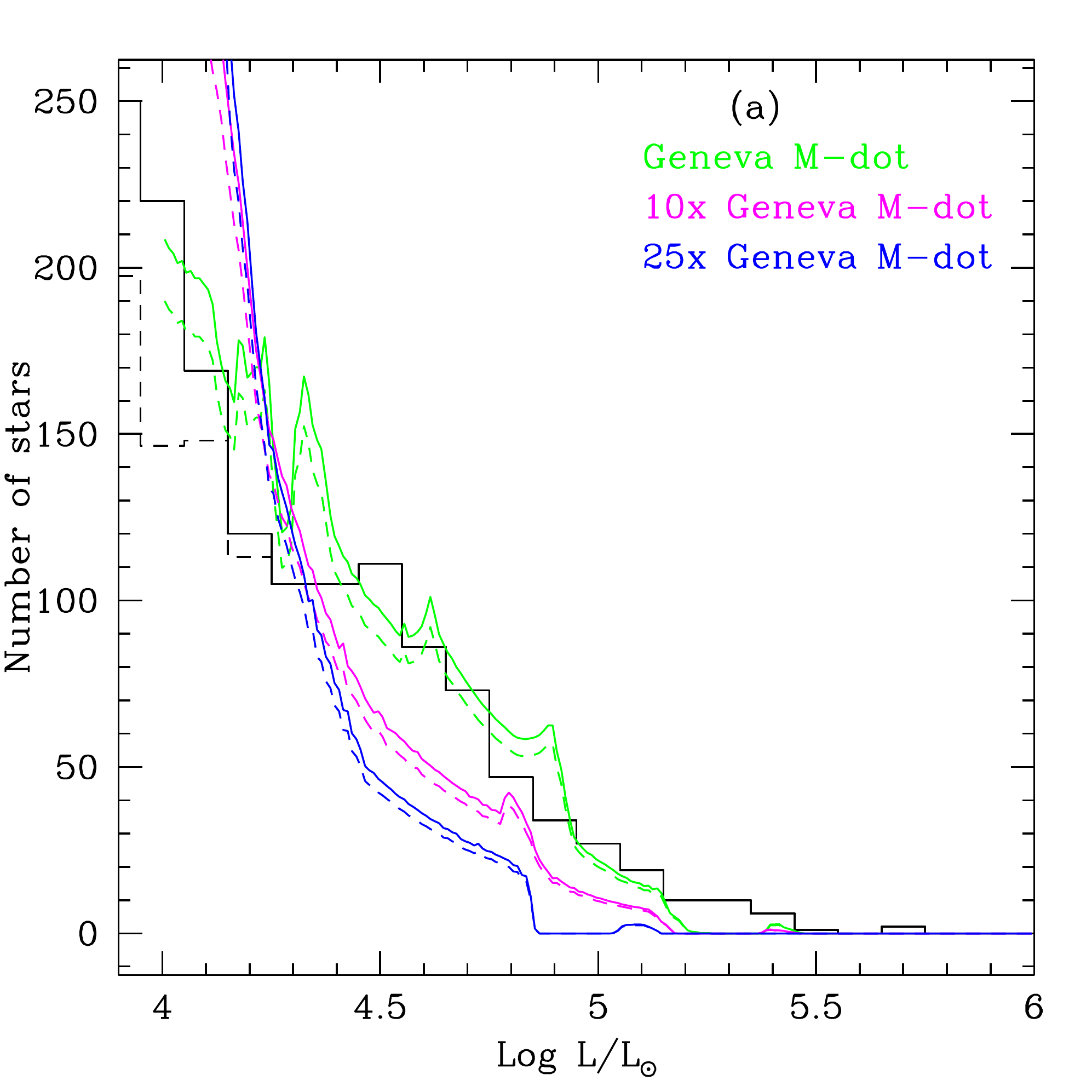}
\plotone{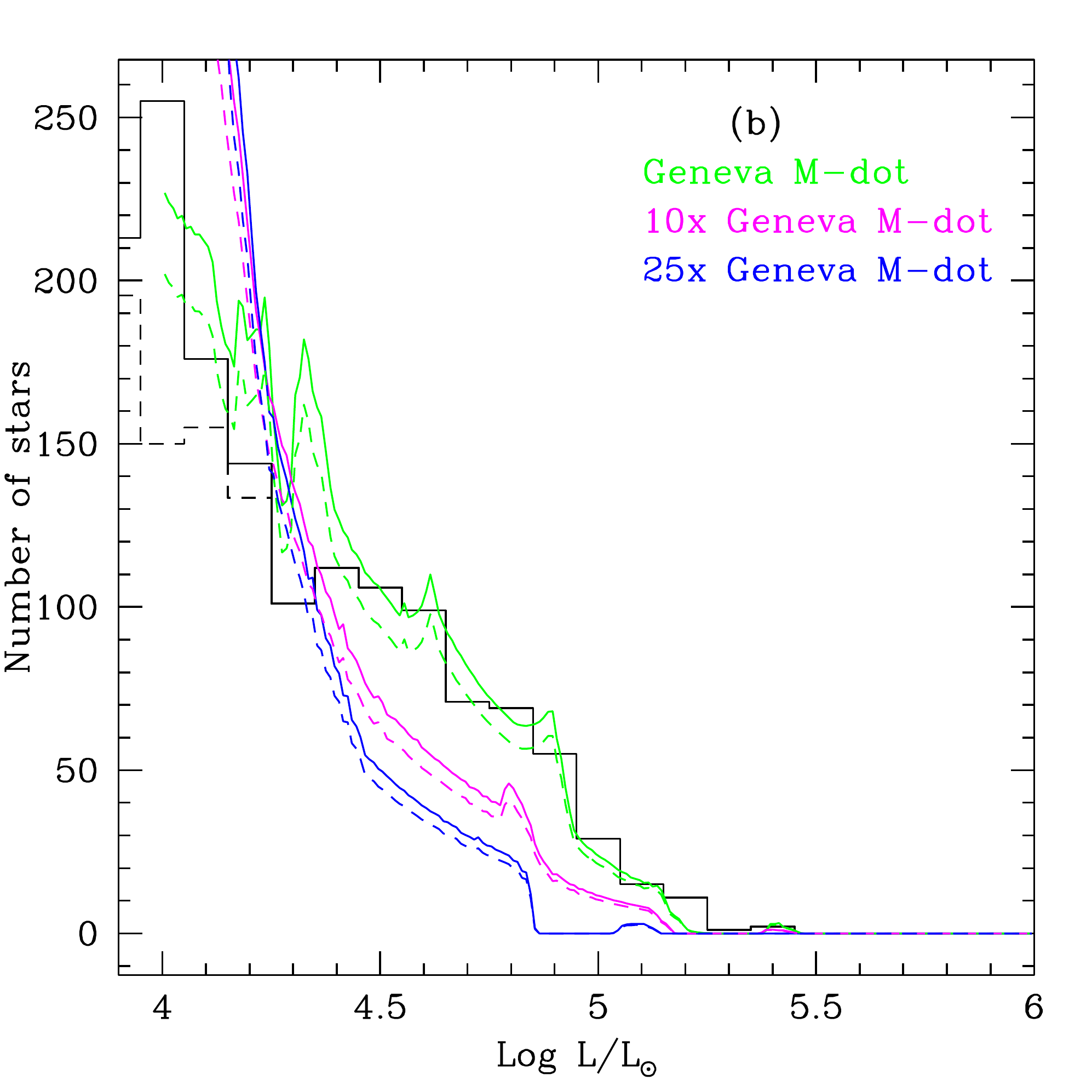}
\plotone{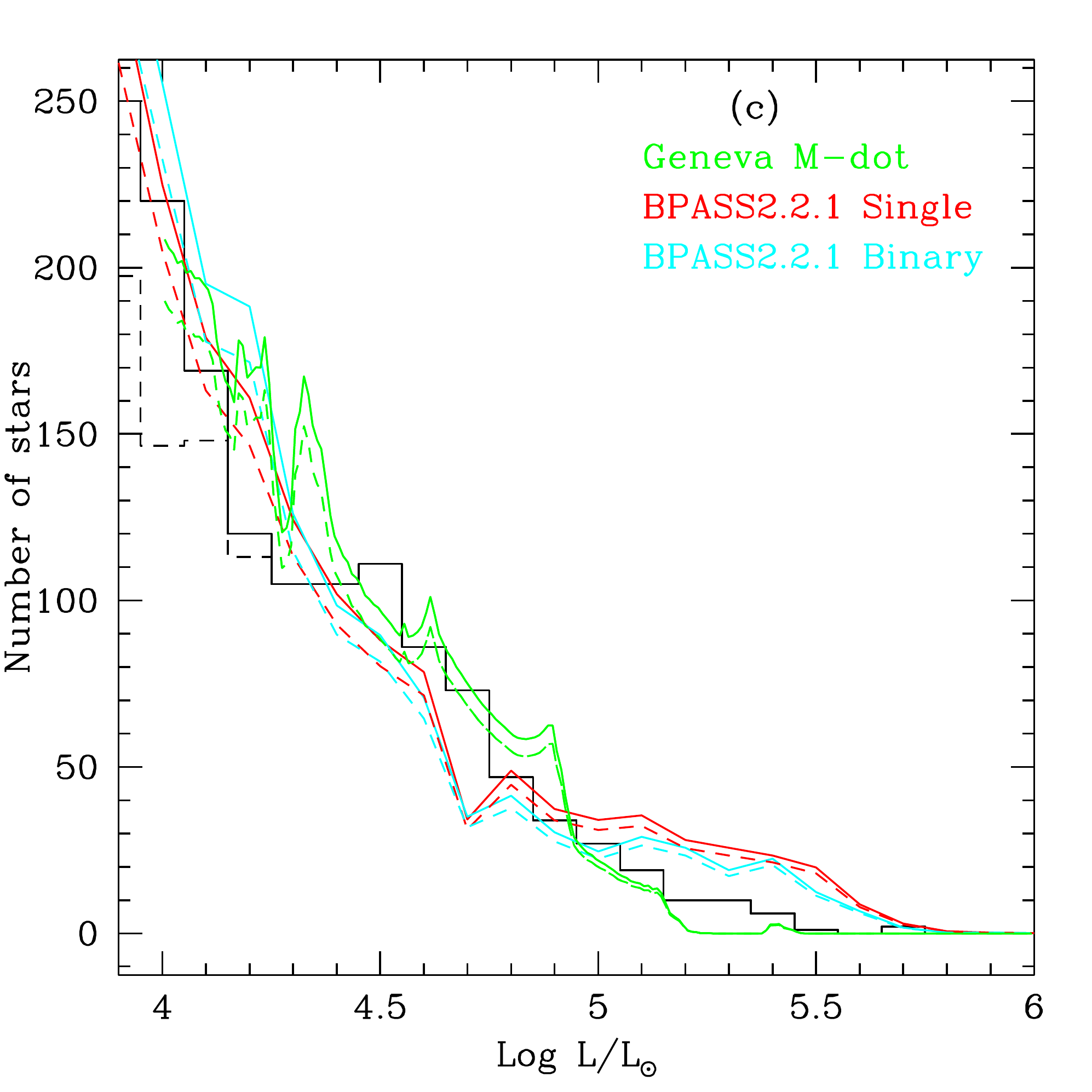}
\caption{\label{fig:lumFunc}Luminosity histogram compared to mass-loss rate model predictions. (a) The figure shows the observed luminosity histogram of M31 RSGs as determined by the UKIRT data.  The black histogram is the combined data from fields A and B in M31, while the dashed shows the effect of the correction of the results from the control field scaled to the same area, and likely is an over-correction.  The {\sc syclist} predictions are shown by three colored curves for different mass-loss rates during the RSG phase.  The currently used Geneva mass-loss rate matches the observations; the 10$\times$ and 25$\times$ enhanced rates predict far too few high luminosity RSGs and too many lower luminosity RSGs for a given number of stars. (b) The same as (a) but using a constant $A_V=1.0$~mag in computing the luminosities. (c) BPASS 2.2.1 model predictions for both single stars (red) and binaries (blue) with the Geneva single-star predictions included in green for comparison.}

\end{figure}

We next compared the luminosity functions given by our sample of M31 RSGs to the {\sc syclist} model predictions. In Figure~\ref{fig:lumFunc}(a) the uncorrected luminosity function is shown by the black histogram; the correction for foreground stars is shown as a dashed line, and recall from the discussion above that this will likely be an {\it over}-correction. We compare this to the three {\sc syclist} predictions: the currently used Geneva mass-loss rate prescriptions (as described in \citealt{Ekstrom2012}), and values enhanced by $10\times$ and $25\times$. There are 1145 stars at luminosities $\log L/L_\odot \ge$ 4.0. The sum of the {\sc syclist} luminosity function is normalized to unity, and we have adjusted the output in accord with the total number of stars and bin size of our histogram. Thus both the shape and the actual values are meaningful in comparing the theoretical predictions with the observations.

We were also curious how much our choice of a luminosity-dependent reddening correction influenced our results.  In Figure~\ref{fig:lumFunc}(b) we show the luminosity function obtained by adopting a constant $A_V=1.0$~mag. We find that our findings are robust: the current Geneva RSG mass-loss prescription still does an excellent job of matching the observations, while the two enhanced ($10\times$ and $25\times$) versions do not. Figure~\ref{fig:lumFunc}(c) includes the BPASS 2.2.1 model predictions for both single and binary stars, showing that these models also match the observations well, except for over predicting the number of high luminosity RSGs. For the BPASS models this is presumably due to the adoption of the \citet{deJager1988} RSG mass-loss rate, without accounting for the supra-Eddington losses that occur for for the highest luminosities.

Our results are quite striking. By comparing the luminosity histograms with the model outputs, we can see that the current prescription for mass-loss rates does an incredibly good job of matching the estimating the number of lower luminosity stars (under $\sim \log L_{\odot} = 4.5$).  In contrast, the $10\times$ and $25\times$ enhanced mass-loss models both underestimate the number of higher luminosity RSGs, and grossly over-estimate the number of lower luminosity RSGs.  This suggests that the RSG mass-loss rates are currently in use are actually a very good approximation to reality both for the Geneva mass-loss rates and BPASS 2.2.1 models.

We can quantitatively describe the agreement between the observed distribution and that of the Geneva models by running a Kolmogorov-Smirnov (KS) test. A comparison between the currently used Geneva mass-loss rates and the observed distribution returns a p-value of 0.97. Comparing our observations to the $10\times$ and $25\times$ enhanced mass-loss models produce p-values of 0.30 and 0.12, respectively. Since the p-value of the current used Geneva mass-loss rates is so close to 1, this objectively shows that they are a good match to observations. KS tests on both the BPASS 2.2.1 single and binary star model models produce p-values of 0.96 suggesting good agreement to the observations as well.

\section{Discussion}
Based on our observations of the luminosities of M31 RSGs, we conclude that the current mass-loss prescriptions in use by the Geneva evolutionary group and described in \citet{Ekstrom2012} are well matched to both our observations and thus what occurs physically. While it is discussed in detail in \citet{Ekstrom2012}, Section 2.6, we briefly go over the mass-loss rate prescriptions used in the models for RSGs here. For RSGs up to 12$M_\odot$, the mass-loss rates are adopted from \citet{Reimers1975,Reimers1977}. For RSGs above 15$M_\odot$, the mass-loss rate depends on the temperature. For stars with log $T_{\rm{eff}} > 3.7$, the rates come from \citet{deJager1988} while for RSGs with log $T_{\rm{eff}} < 3.7$, the rates come from a linear fit to the data from \citet{Sylvester1998} and \citet{vanLoon1999}. For the rotating models (that we use in this paper for comparison), there is an additional correction factor applied as described by \citet{Maeder2000}. However, as we discussed in the Introduction, RSGs are quite unstable and undergo episodic mass-loss rate events with some unknown frequency. In these instances, the external layers of the RSG's envelope exceeds the Eddington luminosity. The Geneva models are not able to model the mass-loss rates in these instances directly using their hydrostatic approaches because the systems are so unstable. Thus, they artificially increase the mass-loss rate of the star by a factor of three whenever the luminosity of any of the layers of the envelope is higher than the Eddington luminosity by a factor of five. This takes into account the differing mass-loss rates caused by episodic instabilities. Based on our observations we find that this method of increasing the mass-loss rates during times of high instability produces a good fit to the observed time-averaged RSG mass-loss rate.

This discovery suggests that the majority of RSGs {\it will not} evolve back across the HRD and become post-RSG YSGs.  We note however that in the upper luminosity range (above $\sim \log L/L_{\odot} = 5.15$, or an initial mass above $20M_{\odot}$), the models of \citet{Ekstrom2012} predict that most of the RSGs will evolve back in the blue. Previous surveys of YSGs in the LMC and M33 found that the Geneva Evolutionary models were able to correctly predict the relative lifetimes of YSGs at varying luminosities \citep{Neugent2012, Drout2012} and our result here lends further credence to the accuracy of the models. From an observational point of view, it is difficult to distinguish the difference between YSGs moving rightward and leftward across the HRD. However, steps are being taken by several authors of this paper to do just that. It has been argued that the stars' rotational velocities will slow down, they may experience periodicity or pulsations, and certain chemical elements will be enhanced. The presence of enhanced pulsations in blue supergiants as post RSG objects was discussed by \citet{Saio2013} and \citet{Georgy2014}. Using TESS data, \citet{Dorn2019} modeled the variability of four YSGs finding that one of them showed drastically different variability from the other three. They hypothesized that this could be a post-RSG object as it is additionally quite dusty. It will be interesting to differentiate between post and pre-RSG YSGs observationally in the future.

Since the RSG lifetime depends upon the mass-loss rate, an additional test will be comparing the number of RSGs to WRs as a function of metallicity.  \citet{Maeder1980} has argued that this ratio should be a sensitive function of metallicity as higher metallicity lends itself to WR production.  The authors of this paper are also currently undertaking a project that will allow us to compare this ratio to model predictions, ideally over a range of  metallicities.

\section{Summary and Conclusions}
As \citet{Geneva} puts it, ``Of the most relevant parameters determining the evolution of massive stars, ... the mass-loss rate during the RSG phase, is the least constrained by the observations or theory." As part of our work on this paper we hoped to remedy this situation by comparing the observed RSG mass-loss rates with those predicted by single-star evolutionary theory. We used new UKIRT data to obtain deep, accurate NIR photometry of stars in two fields in M31. We separated out the RSGs from the foreground stars and AGBs using both {\it Gaia} and color-color cuts down to RSGs of log $L/L_{\odot} > 4.0$. We then calculated the theoretically expected luminosity functions using {\sc syclist} adopting three different time-averaged mass-loss rates. Comparisons with our observed RSG luminosity function show that the actual RSG mass-loss rates are well represented by the Geneva evolutionary models as described by \citet{Ekstrom2012} and the BPASS 2.2.1 single and binary star models as described by \citep{Stanway2018}. Significantly higher (10-25$\times$) mass-loss rates are not consistent with our data.

\acknowledgements
We would like to thank the Arizona TAC for supporting this proposal as well as help from Watson Varricatt in arranging for the UKIRT observations.  The work benefited from a useful correspondence with Ming Yang, and comments on an early version of the manuscript by an anonymous referee. J.J. Eldridge was especially helpful throughout this project by providing BPASS 2.2.1 model predictions and offering comments on the results. This work was supported by the National Science Foundation (NSF) under AST-1612874, and NSF IGERT grant DGE-1258485, as well as by a fellowship from the Alfred P. Sloan Foundation. The Geneva team is thankful to the Swiss National Science Foundation (project number 200020-172505).

\keywords{stars: evolution --- stars: luminosity function, mass function --- stars: mass-loss}
	
\bibliographystyle{apj}
\bibliography{masterbib}

\clearpage
\begin{deluxetable}{l c c c c c r}
\tabletypesize{\scriptsize}
\tablecaption{\label{tab:where} UKIRT Imaging}
\tablewidth{0pt}
\tablehead{
\colhead{Field}
&\colhead{ $\alpha_{2000} $ }
&\colhead{ $\delta_{2000} $ }
&\multicolumn{2}{c}{UT Date of Obs.}
&\colhead{Final Area (deg$^2$)}
&\colhead{\# stars\tablenotemark{a}} \\ \cline{4-5}
&&&
\colhead{Visit 1}
&\colhead{Visit 2}
}
\startdata
A & 0:40:12 & +40:42:00 & 2017 Dec 08 & 2017 Dec 11 & 0.54 & 116,498\\
B &0:45:36 & +42:00:00 & 2017 Dec 30 & 2018 Jan 06 & 0.72 & 132,623\\
Control & 0:35:24 & +41:24:00 &  2017 Dec 12 & 2017 Dec 28 & 0.72 & 11,057\\
\enddata
\tablenotetext{a}{Includes some saturated stars that were removed in the analysis.}
\end{deluxetable}

\clearpage
\begin{deluxetable}{c c c c c c c c c}
\tabletypesize{\scriptsize}
\tablecaption{\label{tab:PhotAB} UKIRT Photometry M31 Fields A and B\tablenotemark{a}}
\tablewidth{0pt}
\tablehead{
\colhead{$\alpha_{2000} $ }
&\colhead{$\delta_{2000}$ }
&\colhead{$J$}
&\colhead{$\sigma_{J}$}
&\colhead{$K_s$}
&\colhead{$\sigma_{K_s}$}
&\colhead{$J-K_s$}
&\colhead{$\sigma_{J-K_s}$}
&\colhead{Mem\tablenotemark{b}}
}
\startdata
00:37:50.46 & +40:59:42.4  & 18.258 &  0.047 & 17.355 &  0.040 &  0.903 &  0.062 & 3\\
00:37:50.62 & +40:54:45.4  & 18.479 &  0.075 & 16.683 &  0.028 &  1.796 & 0.080  &3\\
00:37:50.65 & +40:58:56.8  & 17.805 &  0.042 & 16.508 &  0.024 & 1.297  & 0.048  &3\\
00:37:50.68 & +40:53:17.7  & 18.930 &  0.089 & 17.997 &  0.079 &  0.933 & 0.119  &3\\
00:37:50.82 & +41:06:16.2  & 19.392 &  0.154 & 18.020 &  0.080  & 1.372  & 0.173  & 3\\
\enddata
\tablecomments{Table 2 is published in its entirety in the machine-readable format.
      A portion is shown here for guidance regarding its form and content.}
      \tablenotetext{a}{Stars with $\alpha_{2000}$ less than 00:43 are in Field A; those with larger values are from Field B.} 
\tablenotetext{b}{Membership flag based upon our analysis of the {\it Gaia} DR2 results.  0=member; 1=uncertain; 2=non-member; 3=no {\it Gaia} data available.}
\end{deluxetable}

\clearpage
\begin{deluxetable}{c c c c c c c c c}
\tabletypesize{\scriptsize}
\tablecaption{\label{tab:PhotC} UKIRT Photometry M31 Control Field}
\tablewidth{0pt}
\tablehead{
\colhead{$\alpha_{2000} $ }
&\colhead{$\delta_{2000}$ }
&\colhead{$J$}
&\colhead{$\sigma_{J}$}
&\colhead{$K_s$}
&\colhead{$\sigma_{K_s}$}
&\colhead{$J-K_s$}
&\colhead{$\sigma_{J-K_s}$}
&\colhead{Mem\tablenotemark{a}}
}
\startdata
00:33:01.37 & +41:39:01.0  & 13.935  & 0.002  &13.334  & 0.002 &  0.601  & 0.003 & 2\\
00:33:01.49 & +41:38:28.6   &18.494  & 0.055  &17.793   &0.064  & 0.701 &  0.084 & 3\\
00:33:01.74 & +41:45:13.4  & 15.769  & 0.006  &15.111  & 0.006   &0.658  & 0.008  &2\\
00:33:01.81 & +41:35:33.7  & 16.309  & 0.009  &15.749  & 0.011  & 0.560 &  0.014 & 2\\
00:33:02.02 & +41:38:37.4  & 16.650  & 0.012  &16.124  & 0.015  & 0.526 &  0.019  &2\\
\enddata
\tablecomments{Table 3 is published in its entirety in the machine-readable format.
      A portion is shown here for guidance regarding its form and content. }
\tablecomments{The brightest stars are taken from the 2MASS catalog. See text.}
\tablenotetext{a}{Membership flag based upon our analysis of the {\it Gaia} DR2 results.  0=member; 1=uncertain; 2=non-member; 3=no {\it Gaia} data available.}
\end{deluxetable}

\clearpage
\begin{deluxetable}{c l r}
\tablecaption{\label{tab:FunFacts} Adopted and Derived Relations}
\tablewidth{0pt}
\tablehead{
\colhead{Relation}
&\colhead{Source}
}
\startdata
\sidehead{Adopted Distances:} 
\phantom{MakeSomeSpace} & M31: 760~kpc & 1 \\
& SMC: 59~kpc & 1 \\
\sidehead{Reddening Relations:}
&$A_V=3.1 E(B-V)$ & \nodata \\
&$A_K =0.12 A_V = 0.367 E(B-V)= 0.686 E(J-K)$ & 2 \\
&$E(J-K) = A_V/5.79 = 0.535 E(B-V)$ & 2 \\

\sidehead{Photometric Criteria for RSGs with $K_s\leq 17.0$}
	&$0.87< (J-K_s) \leq 1.0$: $K_0\leq K_s$       & 3  \\
	&$1.0< (J-K_s) \leq 1.3$:       $K_0\leq K_s \leq K_1$ & 3 \\
	&$1.3< (J-K_s) \leq 1.5$:       $K_s \leq 14.4$  & 3 \\
	&$K_0=28.62-13.33(J-K_s)$  & 3,4\\
	&$K_1=31.95-13.33(J-K_s)$  & 3,4\\
	
\sidehead{Adopted Extinction}
	& $K_s>14.5$: $A_V=0.75$ & 3 \\
	& $K_s<14.5$: $A_V=0.75-1.26(K_s-14.5)$ & 3 \\

\sidehead{Conversion from 2MASS ($J, K_s$) to Standard System ($J, K$):}
&$K=K_s + 0.044$ & 5\\
&$J-K = (J-K_s+0.011)/0.972$ & 5\\

\sidehead{Conversion to Physical Properties (Valid for 3500-4500 K):}
&$T_{\rm eff} = 5643.5 - 1807.1 (J-K)_0$ & 3\\
&${\rm BC}_K = 5.567 - 0.7569 \times T_{\rm eff}/1000$ & 3\\
&$K_0 = K - A_K$ & \nodata \\
&$M_{\rm bol} = K_0 + {\rm BC}_K - 24.40$ & \nodata \\
&$\log L/L_\odot =(M_{\rm bol}-4.75)/-2.5$ & \nodata \\

\enddata
\tablerefs{1--\citealt{vandenbergh2000}; 2--\citealt{Schlegel1998}; 3--This paper; 4--\citealt{Cioni2006a}; 5--\citealt{Carpenter}}
\end{deluxetable}

\end{document}